\documentclass[aps,floats,twocolumn,epsf,prl,showpacs,nofootinbib]{revtex4-1}
\usepackage{graphicx}
\usepackage{epstopdf}
\usepackage{amsmath}
\usepackage{amssymb}
\usepackage[colorlinks=true,urlcolor=blue,citecolor=blue,linkcolor=blue,breaklinks=true]{hyperref}
\usepackage{color}
\usepackage{float}
\usepackage{hyperref}
\usepackage[capitalise]{cleveref}
\usepackage[normalem]{ulem}
\usepackage[dvipsnames]{xcolor}

\usepackage{bm}
\usepackage{enumitem}

\newcommand{\ii}{\mathrm{i}}
\newcommand{\dd}{\mathrm{d}}
\newcommand{\e}{\mathrm{e}}

\newcommand{\expec}[1]{\langle {#1} \rangle}
\newcommand{\vk}{{\bm{k}}}

\newcommand{\vQ}{{\bm{Q}}}

\newcommand{\operator}[1]{\hat{#1}}
\newcommand{\creation}[1]{\operator{c}_{#1}^\dagger}

\newcommand{\annihilation}[1]{\operator{c}_{#1}^{\phantom{\dagger}}}
\newcommand{\occopp}{\operator{n}}
\newcommand{\V}{\mathcal{V}}

\newcommand{\dga}{\textrm{D}\Gamma\textrm{A}}

\usepackage{orcidlink}

\begin{document}

\title{Fermi and Luttinger arcs: two concepts, realized on one surface}

 \author{Paul Worm\,\orcidlink{0000-0003-2575-5058}}
\thanks{These authors contributed equally}
 \affiliation{Institute of Solid State Physics, TU Wien, 1040 Vienna, Austria}
 \author{Matthias Reitner\,\orcidlink{0000-0002-2529-0847}}
\thanks{These authors contributed equally}
 \affiliation{Institute of Solid State Physics, TU Wien, 1040 Vienna, Austria}
 \author{Karsten Held\,\orcidlink{0000-0001-5984-8549}}
 \affiliation{Institute of Solid State Physics, TU Wien, 1040 Vienna, Austria}
  \author{Alessandro Toschi\,\orcidlink{0000-0001-5669-3377}}
 \affiliation{Institute of Solid State Physics, TU Wien, 1040 Vienna, Austria}

\date{\today}

\begin{abstract}
We present an analytically solvable model for correlated electrons, which is able to capture the major Fermi surface modifications occurring in both hole- and electron-doped cuprates as a function of doping. The proposed Hamiltonian qualitatively reproduces the results of numerically demanding many-body calculations, here obtained using the dynamical vertex approximation. 
Our analytical theory provides a transparent description of a precise mechanism, capable to drive the formation of disconnected segments along the Fermi surface (the highly debated ``Fermi arcs'') as well as of the opening of a pseudogap at hole- and electron-doping. This occurs through a specific mechanism: The electronic states on the Fermi arcs remain intact, while the Fermi surface part where the gap opens transforms into a Luttinger arc.
 \end{abstract}

\maketitle

\noindent
{\sl Introduction}---The emergence of isolated arcs on the Fermi 
surface  (FS) of strongly correlated systems, such as those characterizing the famous pseudogap regime~\cite{Timusk99,Norman1998} of high-temperature superconducting cuprates~\cite{Keimer15}, has been the object of an incessant and quite vivid theoretical debate in the last thirty-five years.

In the specific case of the cuprates, the presence of Fermi arcs is not only suggested by the evolution of the ARPES spectra as a function of temperature and doping~\cite{Vishik2018,Reber2012,Kunisada2020,Shen05,Kordyuk2015,Lee2007,Hashimoto2014,Vishik_2010,Vishik2012,Hashimoto2015,Rui2011} but also by the persistence of clear Fermi liquid features in the qualitative behavior of transport experiments performed in the whole normal phase of these compounds~\cite{Barisic2013,Barisic2019}. 

At the same time, the emergence of Fermi arcs remains quite puzzling from a theoretical point of view:  On the one hand, they would be forbidden in conventional band theory~\cite{Wan2011,Schrunk2022}(where bands are continuous and Fermi surfaces closed)
and, on the other hand, the complexity associated with the explicit inclusion of electronic correlations hinders the identification of a transparent mechanism explaining their formation.

In the latter respect, the quest for explaining the enigmatic properties of the celebrated pseudogap phase in the cuprates has led many groups to perform a huge number of theoretical calculations, ranging from phenomenological modeling of the pseudogap self-energy \cite{Sadovskii2005,Yang2006,Yamaji2011,Rice2012,Sakai2016,Robinson2019,Fabrizio2022, WormThesis} to advanced quantum many-body calculations (mostly of hole- or electron-doped Hubbard/Hubbard-like models)~\cite{Sakai2009,Qin2022,schafer2021foodprints,krien2022pseudogap,Tremblay2006,VanLoon2018,Wu2018,Wagner2023}. In several cases, the theoretical results have provided spectral functions and transport properties compatible with the experimental observations in the cuprates \cite{Benfatto2000, bonetti2020chargedrop}. Further, the application of the fluctuations diagnostics approach \cite{gunnarsson2015fluctuation,Wu2017diag,rohringer2020spectra,krien2020fluctuation,schafer2021fluctuations,Wu2022,dong2022supercon} has highlighted the predominant role played by strong spin fluctuations \cite{Scalapino2012,Huscroft2001,Senechal2004,kyung2006,Macridin2006} in driving the outcomes of quantum many-body calculations in the interaction and doping regimes relevant to cuprates \cite{Lee2003}. 

However, neither the phenomenological approaches nor the most advanced numerical studies have so far provided an intuitive and physically intelligible picture of the observed phenomena.

In this paper we approach the problem from a different perspective: We introduce an {\sl exactly solvable} many-electron model Hamiltonian on a two-dimensional (2D) lattice. This Hamiltonian is able to describe---depending on the parameter regimes considered---the formation of Fermi arcs in different sectors of the FS and to capture the qualitative differences in the spectral properties between electron- and hole-doped compounds, including the pseudogap properties.
It is worth underlining already here that, while the conception of our Hamiltonian is---to a given extent---inspired by the Hatsugai-Kohmoto model (HKM)~\cite{HK1992}, which has recently attracted an increasing interest both for its physical content and its formal properties~\cite{Yeo2019entropy,Phillips2020,Yang2021exactly,Zhong2022,Li2022sc,Mai2023,Wang2023,Wysokinski2023,Zhao2023,Zhao2023Friedel,Yang2023bose,Skolimowski2024,setty2023electronic,setty2023symmetry}, its key ingredient presents a fundamental difference: The electronic interaction of our model is {\sl not} purely diagonal in momentum space, as it is in the HKM, but couples all momenta differing by the transfer momentum of  $\vQ=(\pi,\pi)$ associated to antiferromagnetic (AF) spin fluctuations. As we demonstrate in the following, this feature enables our model, differently from the HKM,  to capture the fundamental momentum-dependent modifications of the FS occurring in 2D correlated electron systems as a function of hole/electron doping, in reasonable agreement to those we obtain by means of cutting-edge quantum many-body schemes, such as the dynamical vertex approximation (D$\Gamma$A)~\cite{Toschi2007,Katanin2009,WormThesis}.  
In this context, the invaluable insight gained by mastering an analytically solvable model provides a new route to clarify the hotly debated questions of the emergence of Fermi arcs, of the mechanisms controlling their evolution, and, consequently, of the intrinsic nature of the pseudogap phase.

\noindent
{\sl Model Hamiltonian}---We introduce here our model Hamiltonian for a square lattice, which (i) allows for a general exact analytical solution and (ii) displays Fermi arcs, defined as disconnected parts of the FS. 
It reads: 
\begin{equation} 
    H = \sum_{\vk \sigma } \left[(\epsilon_\vk-\mu)\occopp_{\vk \sigma } + \frac{\V}{2} \occopp_{\vk \sigma} \occopp_{\vk+\vQ -\sigma }\right],
\label{eq:our_model}
\end{equation}
where $\mu$ is the chemical potential, $\occopp_{\vk \sigma } \!=\! \creation{\vk \sigma } \annihilation{\vk \sigma}$ the occupation operator, $\V$ the interaction strength that couples momenta $\vk$ and $\vk+\vQ$,  with $\vQ \!=\! (\pi,\pi)$ describing the characteristic momentum of the coupling. For the dispersion relation $\epsilon_\vk$, we consider:

\begin{equation}
\begin{split}
        \epsilon_\vk = &-2t (\cos(k_x)+\cos(k_y)) 
        - 4t' \cos(k_x) \cos(k_y)\\
        &-2t'' (\cos(2k_x)+\cos(2k_y)),
\end{split}
\end{equation}
where we use $t' = -0.2$ and $t'' = 0.1$ (while $t = 1$ sets the unit of energy) consistent with typical values used in low-energy tight-binding models for cuprates \cite{Nicoletti2010a,WormThesis}. See SM~\cite{supplemental} for a brief discussion regarding the role of the sign of the hopping parameters.

\noindent
{\sl Relation to the  Hatsugai-Kohmoto model}---As mentioned in the introduction, our Hamiltonian can be linked to the Hatsugai-Kohmoto model (HKM) \cite{HK1992}, which can be retrieved by setting  $\vQ = 0$  in Eq.~(\ref{eq:our_model}) and also allows for an exact solution. 
We recall here that the HKM can be mapped to a set of decoupled atomic limit models [$H_{AL} = U \occopp_\uparrow  \occopp_\downarrow - \mu_{\rm{eff}} (\occopp_\uparrow \! + \!\occopp_\downarrow)$] in the absence of a magnetic field, one for each $\vk$ point, whose difference is only in the value of their effective chemical potentials $\mu_{\rm{eff}} = \mu - \epsilon_\vk$. This property, which renders the model solvable, also allows for a description of the appearance of Luttinger zeros at the Fermi energy and of Mott-like insulating phases \cite{HK1992,Continentino1994,Vitoriano2000,Yeo2019entropy}. At the same time, we note that the complete decoupling of the HKM into decoupled  $H_{AL}$ models, severely restricts the possibility to describe second-order phase transitions. Further, recent work \cite{guerci2024} has pointed out possible intrinsic inconsistencies in the linear response expressions of the HKM.

The crucial difference of the HKM  w.r.t.~our model arises from the nonzero value of $\vQ =(\pi,\pi)$  in Eq.~(\ref{eq:our_model}). This captures the coupling of electrons at different momenta (absent in the momentum-diagonal HKM), mediated by AF fluctuations, eventually allowing the emergence of momentum-selective behaviors. To a given extent,  Eq.~(\ref{eq:our_model})  can be viewed as a set of atomic limit models in momentum space in the presence of {\sl a magnetic field} $h_{\vk} = (\epsilon_\vk-\epsilon_{\vk+\vQ})/2$ and chemical potential $\tilde{\mu}_{\vk} = \mu - (\epsilon_\vk+\epsilon_{\vk+\vQ})/2$ acting on two states of opposite spin at different momenta (i.e., ${\vk \! \uparrow}$, ${\vk\!+\!\vQ\! \downarrow}$) (see also SM Fig.~S3 \cite{supplemental}).

\begin{figure}[tb]
	\includegraphics[width=1.\columnwidth]{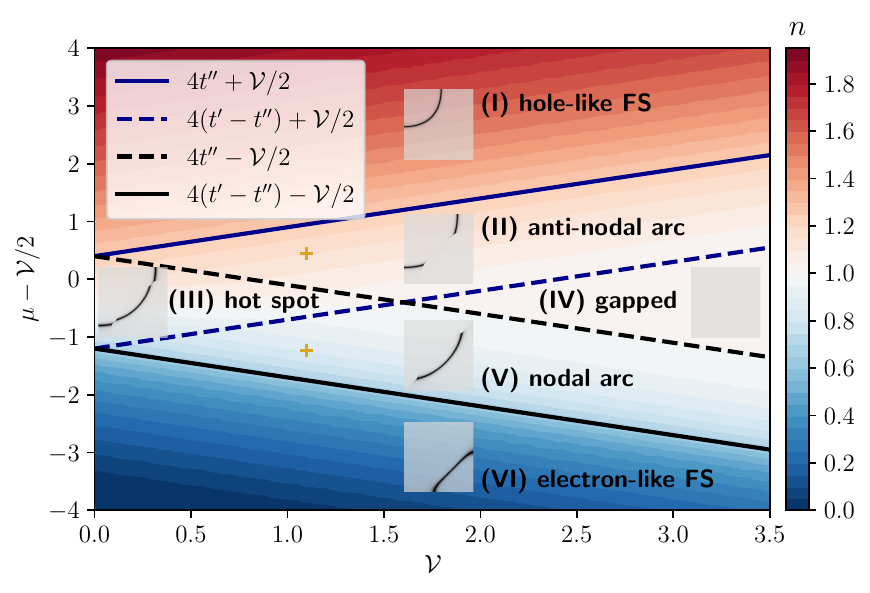}
	\caption{Different phases of disconnected Fermi surfaces realized in the model Hamiltonian for $T\to0$ ($\beta=1/T=10^6/t$). The color scale in the background displays the electron density $n$ as a function of the chemical potential $\mu$ and interaction strength $\V$. The black and blue solid and dashed lines, which depend on the interaction and the dispersion $\epsilon_\vk$ ($t=1$, $t^{\prime}=-0.2$, $t^{\prime \prime}=0.1$) divide the parameter space into six regions of qualitatively different spectral functions: (I) hole-like Fermi surface, (II) anti-nodal Fermi arc, (III) pseudogap on the hotspots, (IV) completely gapped, (V) nodal Fermi arc, and (VI) electron-like Fermi surface. An exemplary momentum-resolved spectral function of each region is displayed in the small insets. The gold ``$+$" symbols indicate the values of $\V$ and $\mu$ used in Fig.~\ref{fig:fermi_surface} and \ref{fig:arc_dispersion}.}
	\label{fig:model_phasediag}
\end{figure}

\noindent
{\sl Analytic solution}---To analyze the evolution of the FS, and in particular the possible emergence of disconnected sections, we start by computing the single-particle Green's function as a function of the Matsubara frequency $\nu_n$: 
$
    G_{\vk \sigma}(\ii \nu_n) \equiv  - \int_{0}^{\beta}\! d\tau \, \expec{\annihilation{\vk \sigma}(\tau) \creation{\vk \sigma}(0)}\,  \e^{\ii \nu_n \tau }.
$

For the model proposed in \cref{eq:our_model}, $G_{\vk \sigma}$  can be analytically computed directly on the real-frequency axis
\begin{equation}
    G_{\vk}( \omega) = \frac{1-n_{\vk+\vQ}}{\omega + \mu - \epsilon_\vk+ \ii 0^+} + \frac{n_{\vk+\vQ}}{\omega + \mu - \epsilon_\vk - \V+ \ii 0^+},
\label{eq:G_our_model}
\end{equation}
where $n_{\vk+\vQ} = \expec{\occopp_{\vk+\vQ \sigma}}$ (assuming no magnetic order) can be calculated according to Eq.~(\ref{eq:n_our_model}) below.
We note that the presence of $n_{\vk+\vQ}$ (instead of $n_{\vk}$ as in the HKM), allows here for the coupling of different momenta, and hence for the possible emergence of momentum-selective features like Fermi arcs.
Specifically, there are three different (momentum-dependent) possibilities for the Green's function: (i) for all momenta $\vk$ at which $n_{\vk+\vQ} = 0$, we have the non-interacting Green's function; (ii) for all $\vk$ at which $n_{\vk+\vQ} = 1$ only the second term in the r.h.s.~of Eq.~(\ref{eq:G_our_model}) is nonzero,  yielding the non-interacting solution but with a chemical potential shifted by $\V$; and (iii) for all $\vk$ at which $0 \! <n_{\vk+\vQ} <\!  1$, both branches of $G$ in Eq.~(\ref{eq:G_our_model})  are active and the spectrum splits into two peaks with a gap in between.

For the four many-body states of each pair ${\vk \! \uparrow}$, ${\vk\!+\!\vQ\! \downarrow}$ (see SM~\cite{supplemental}),
we can calculate  
\begin{equation}
    n_{\vk+\vQ}= \frac{\e^{-\beta \xi_{\vk+\vQ}} + \e^{-\beta (\xi_{\vk}+\xi_{\vk+\vQ} + \V)} }{1+\e^{-\beta \xi_\vk} + \e^{-\beta  \xi_{\vk+\vQ}} + \e^{-\beta (\xi_{\vk}+\xi_{\vk+\vQ} + \V)}},
\label{eq:n_our_model}
\end{equation} 
where $\beta = 1/T$ is the inverse temperature, and $\xi_\vk = \epsilon_{\vk} - \mu$.
This yields the evolution of the FS as a function of the chemical potential $\mu$ and the interaction strength $\V$ in the low-$T$ limit ($\beta \! \rightarrow \infty$) shown in \cref{fig:model_phasediag}. Here, the color plot displays the corresponding filling, and the lines separate regions of qualitatively different Fermi surfaces. A representative quadrant of the corresponding FS is shown as inset in each region.

This phase diagram can be understood further. Let us start with the arguably most interesting hole-doped case (V).  For the part of the FS, that lies inside the AF zone boundary (AFZB),
i.e. inside the AFZB lines  $\pm k_x  \pm k_y = \pi$, 
$\vk$
couples via $\vQ$  to empty states with $n_{\vk+\vQ} \!=\! 0$. Thus we obtain the non-interacting spectrum $A(\omega,\vk)\!=\!\delta(\omega-\xi_\vk)$; the FS remains intact. Contrary, the part around the antinode [around $\vk =(\pi,0)$], which is outside of the AFZB, is coupled to filled states with $n_{\vk+\vQ}\! =\! 1$. Consequently, the spectrum $A(\omega,\vk)\!=\!\delta(\omega-\xi_\vk-\V)$ is shifted to energies above the Fermi level.
Hence, there is no FS for such momenta. 
Instead, a Luttinger surface (LS) appears \cite{Dzyaloshinskii2003}, defined 
by $G_\vk(\omega=0) \!=\! 0$ (i.e., with both real and imaginary part vanishing). Resolving 
Eq.~(\ref{eq:G_our_model}) for this condition yields 
$\xi_\vk =\epsilon_{\vk} - \mu\!=\!(n_{\vk+\vQ}-1) \V$,
which for $\beta \to \infty$ becomes fulfilled close to the AFZB.

The result is a disconnected FS, or more precisely: a transition of a FS into a LS at the AFZB. This insight of an exactly solvable model is crucial to understand Fermi arcs: It shows that Fermi surfaces do not need to be connected, but can display a direct transition into Luttinger arcs, which are invisible to ARPES experiments. {The underlying combined surface (LS and FS) defined by $\Re G_\vk(\omega=0)\!=\! 0$ 
remains continuous and connected. However,  it splits into Fermi arcs 
[poles of $\Im G_\vk(\omega=0)$] and Luttinger arcs [zeros of $\Im G_\vk(\omega=0)$]}, see also SM~\cite{supplemental}.
 The location of the Fermi arcs in our model is qualitatively consistent with ARPES experiments in cuprates \cite{Vishik2018,Reber2012,Kunisada2020,Shen05,Kordyuk2015,Lee2007,Hashimoto2014,Vishik_2010,Vishik2012,Hashimoto2015,Rui2011}.
 Further, the evolution of the FS/LS as a function of doping is directly reflected in the corresponding low-$T$ behavior of thermodynamic  properties, such as the electronic specific heat for $T\rightarrow 0$ (see~SM~\cite{supplemental}).
 
\begin{figure}[tb]
	\includegraphics[width=0.95\columnwidth]{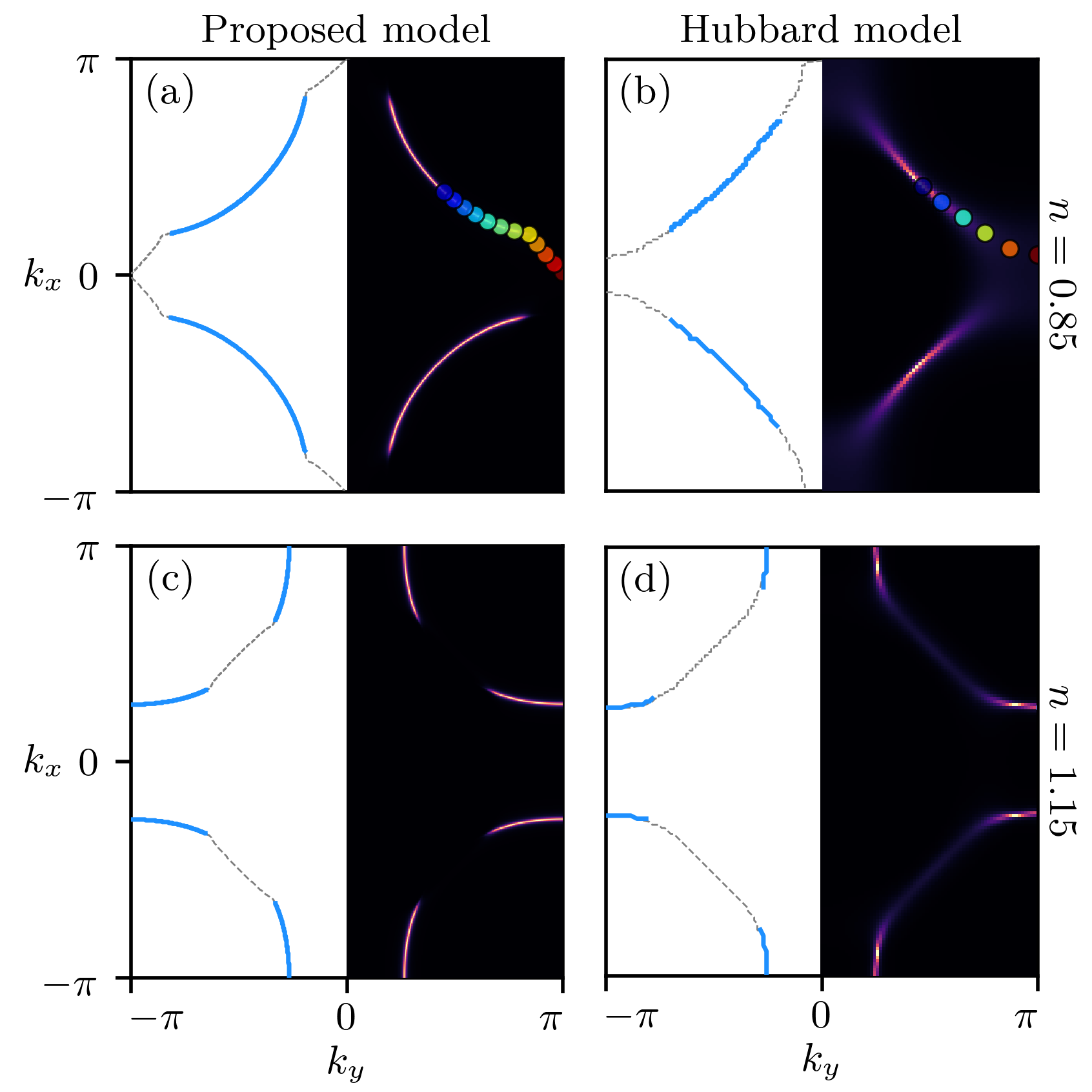}
	\caption{Fermi surface (blue line) and Luttinger surfaces (gray line) computed for the proposed model at $\V=1.1t$ (left column) and, by means of $\dga$, for the Hubbard model at $U=8t$ (right column) at hole- (top row) and electron-doping (bottom row) at $\beta = 12.5/t$. The left half of the Brillouin zone displays the Fermi surface (blue line) and Luttinger surface (gray dashed), while the right half shows the spectral function at zero frequency ($A(\omega=0)$). Colored dots in the top row mark locations for the cuts in Fig.~\ref{fig:arc_dispersion}.}
	\label{fig:fermi_surface}
\end{figure}

For the electron-doped case (II), the scenario gets reversed w.r.t.~to the AFZB, see panels (a) and (c) of  Fig.~\ref{fig:fermi_surface}. This again agrees with the experimental observation in cuprates \cite{Armitage01,Armitage2002,Armitage2010}.

It is also worth stressing here that 
the close proximity of Luttinger and Fermi {arcs}, which characterizes the pseudogap regimes (II) and (V), may lead one to suspect a mismatch in the Luttinger count \cite{Luttinger1960} $n \!=\! \sum_{\vk \sigma} \Theta(\Re G_\vk(\omega\!=\!0))$, similar to the one reported in Mott insulators~\cite{Stanescu2007, Rosch2007}. In fact, for momenta close to the Luttinger surface one does find that $\frac{1}{2\pi}\int^\infty_{-\infty}\dd\nu\, \e^{\ii \nu 0^+} G_\vk(\ii\nu) \frac{\partial\Sigma_\vk(\ii \nu)}{\partial\ii \nu} \neq 0$ \cite{Altshuler1998,skolimowski2022}, being $\Sigma_\vk(\ii \nu)$  the corresponding self-energy. However, for $T \to 0$, where Luttinger's theorem applies~\cite{Luttinger1960, abrikosov2012methods}, such violations remain confined on one-dimensional paths in the Brillouin zone (the AFZB) and hence vanish in the corresponding momentum integration.

Let us also briefly discuss the other physical regimes shown in Fig.~\ref{fig:model_phasediag}. In particular, since the Fermi surface becomes disconnected only at the AF zone boundary, all values of $\mu$, $\V$ for which $\xi_\vk=0$ and $\xi_\vk+\V=0$ do not cross the AF zone boundary, will show only a regular continuous hole- (I) or electron-like (VI) Fermi surface or a completely gapped half-filled system (IV). From this restriction we can deduce the two regions: (II), $4(t^{\prime}-t^{\prime \prime} )+\V < \mu  < 4t^{\prime \prime}+\V$, and (V),  $4(t^{\prime}-t^{\prime \prime} ) < \mu  < 4t^{\prime \prime}$, where the (anti-)nodal arcs are present. When the two regions intersect, both arcs are visible (III), similar to the so-called ``hotspots'' observed in experiment \cite{Armitage2002,Matsui2007}.

\noindent
{\sl Connection to the Hubbard model}---After illustrating the mechanisms through which \cref{eq:G_our_model} displays Fermi arcs, an important question to be addressed is how our findings connect to the physics described by the microscopical Hamiltonian typically adopted for describing the low-energy properties of cuprates, namely the two-dimensional Hubbard model (HM):
\begin{equation}
    H = \sum_{\vk \sigma } (\epsilon_\vk-\mu)\occopp_{\vk \sigma } + \frac{U}{2}\sum_{i \sigma} \occopp_{i \sigma} \occopp_{i -\sigma},
\label{eq:Hubbard_model}
\end{equation}
where $\epsilon_\vk$ is the same dispersion as in Eq.~(\ref{eq:our_model})  and $U \!= \! 8t$ is the strength of the on-site electrostatic repulsion. It is well known that any reliable calculation of the spectral properties of the HM requires advanced numerical many-body approaches \cite{Qin2022}. Here, we resort to the dynamical vertex approximation ($\dga$) \cite{Toschi2007,Katanin2009, WormThesis}, which is a diagrammatic extension \cite{rohringer2018diagrammatic,kitatani2022} of the dynamical mean-field theory \cite{georges1996dynamical,Vollhardt2012,Georges1992}, capable of treating non-local spatial correlations \cite{rohringer2016impact} even in the intermediate-to-strong coupling regimes of cuprate physics \cite{Pruschke1996}. In fact,  while not exact, $\dga$ compares well to numerically exact quantum Monte Carlo, where such solutions are available \cite{schafer2021foodprints} and was recently also successfully applied to the study of nickelate superconductors \cite{Held2022}. 

\begin{figure}[tb]
	\includegraphics[width=1\columnwidth]{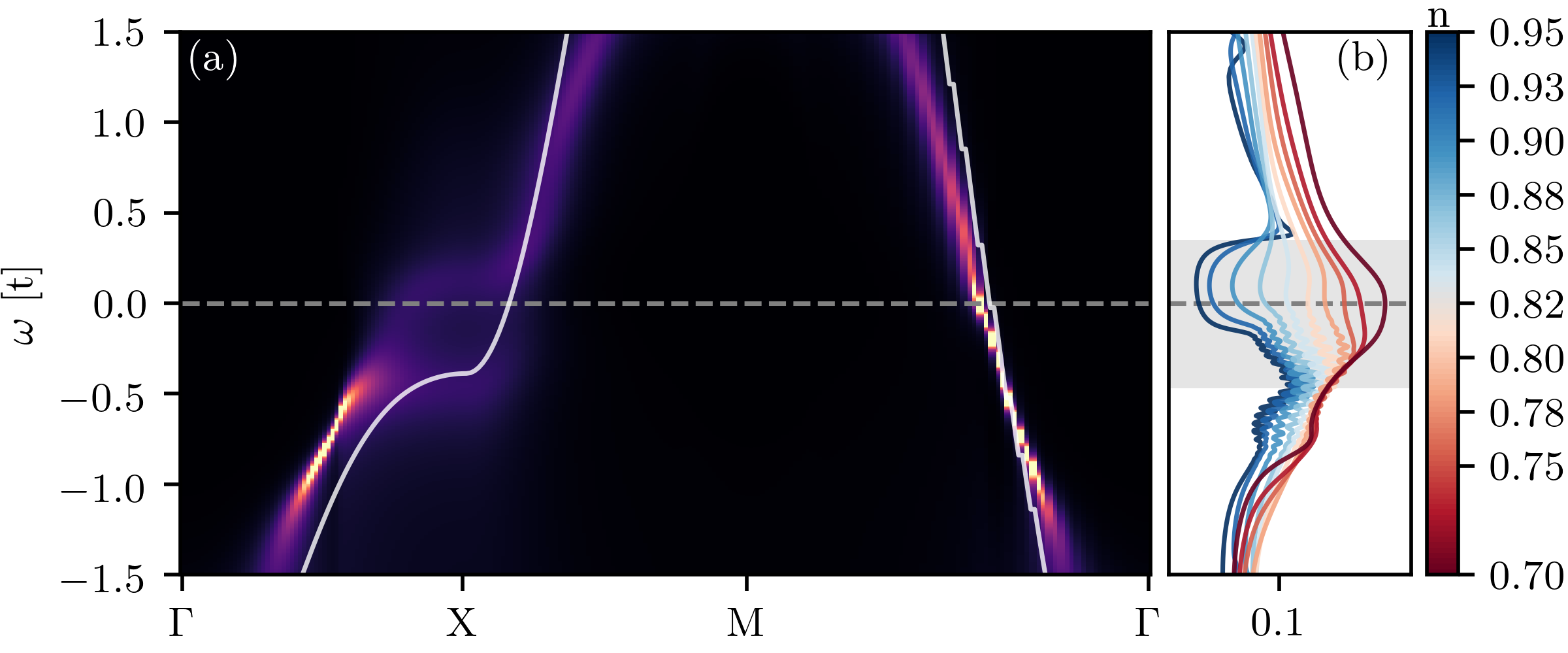}
	\caption{(a) Spectral function $A(\omega,k)$ along a high-symmetry path $\Gamma = (0,0)$, X$ = (\pi,0)$ and M$ = (\pi,\pi)$, for the analytically continued $\dga$ solution of the Hubbard model at $n = 0.85$ and $U=8t$. The white line displays the non-interacting band dispersion for the same filling. Visible is the dichotomy of the spectrum between the antinode, which is gapped, and the node which is not. This is directly reflected in the dip at the Fermi energy of the momentum-integrated spectral function $A(\omega)$, displayed in panel (b) for a range of dopings. As expected and observed in cuprates~\cite{Shen05} the pseudogap is stronger at lower doping and we observe no suppression for $n \leq 0.7$.}
	\label{fig:pseudogap_doping_dependence}
\end{figure}

We summarize our D$\Gamma$A results for the FS of the HM, calculated at $\beta = 12.5/t$ (for lower temperatures and computational details see SM~\cite{supplemental}), both at 
hole- and electron doping in the right panels of  
Fig.~\ref{fig:fermi_surface}.  Here, clear pseudogap features are visible in terms of a significant suppression of the Fermi-level spectral intensity at the antinode (node) for the hole- (electron-) doped case, while the FS around the node (antinode) remains quite sharp and well-defined. The overall behavior can be more precisely understood by looking at the full frequency dependence of the spectral function $A(\omega, \vk)$ along a high-symmetry path across the BZ, for a specific case ($n=0.85$, s.~Fig.~\ref{fig:pseudogap_doping_dependence}a). Evidently, this interaction-driven momentum-selective behavior at the Fermi energy well resembles the one observed in our proposed model \cref{eq:our_model}. In particular, both our model and the $\dga$ solution of the HM  yield the same evolution of Fermi- (blue lines) in Luttinger arcs (gray dashed), as displayed in Fig.~\ref{fig:fermi_surface}. 
At the same time, we note that the Fermi arcs computed in $\dga$ do not end at the AFZB, but are more gradually smeared out, directly reflecting the momentum evolution of the (imaginary part of) $\dga$-self-energy, which is smallest at the node (antinode) and increases along the arc, thus resulting in a broader spectrum. This feature is directly reflected in the doping evolution of the $\vk$-integrated spectral function $A(\omega)$, shown in Fig.~\ref{fig:pseudogap_doping_dependence}b) for the hole-doped case.

 \begin{figure}[tb]
	\includegraphics[width=0.95\columnwidth]{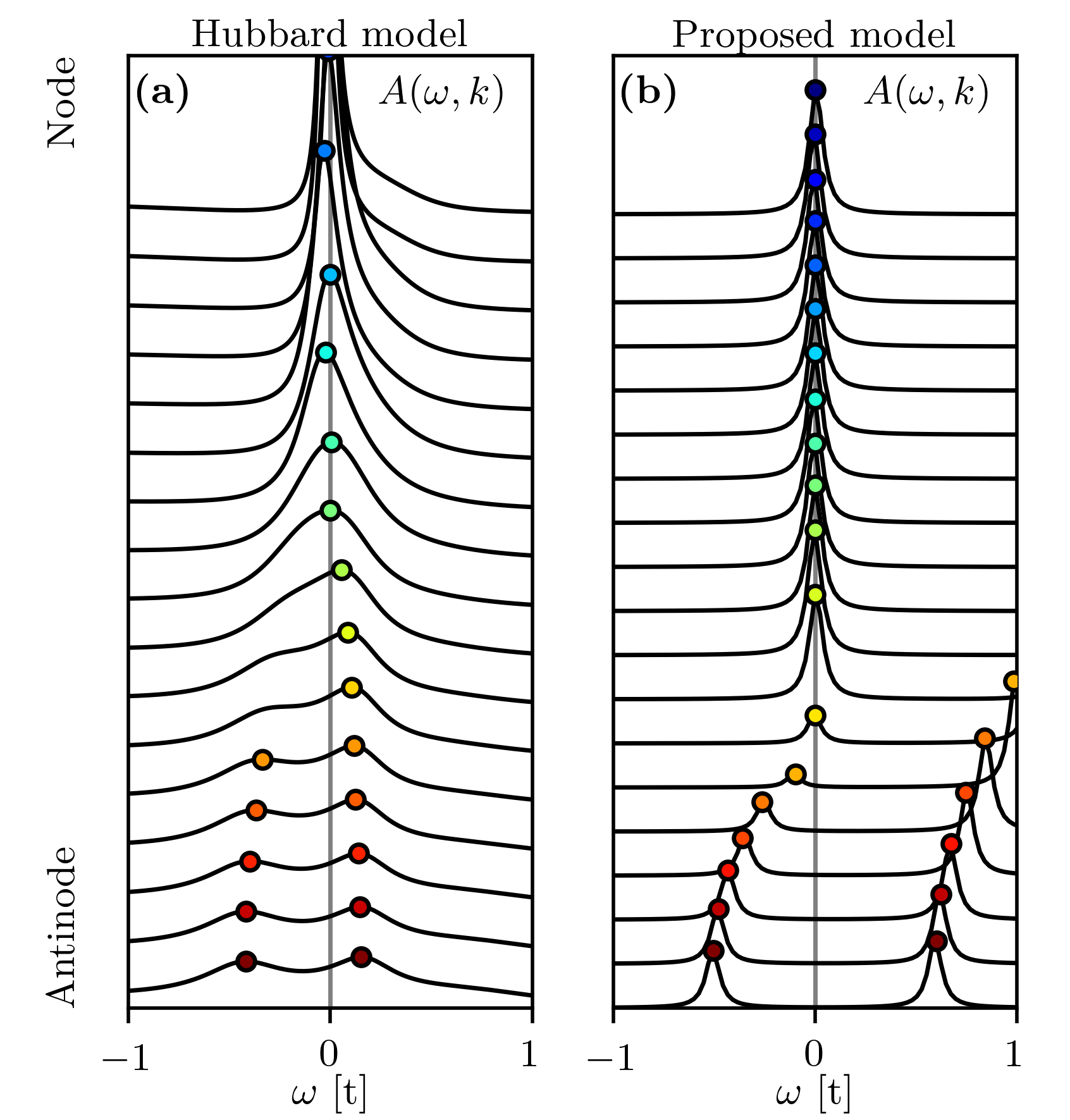}
	\caption{Structure of the spectral function along the Fermi-Luttinger surface, for the momenta indicated by the same color in Fig.~\ref{fig:fermi_surface}. Both $\dga$ and the model are shown for the same filling $n = 0.85$ and inverse temperature $\beta = 12.5/t$.
 Left: Hubbard model using D$\Gamma$A at $U=8t$. Right: exact solution of the model Eq.~\ref{eq:our_model} at $\V=1.1 t$. }
	\label{fig:arc_dispersion}
\end{figure}

In order to visualize the full frequency dependence of the spectrum along the arc, we plot $A(\omega,\textbf{k})$ in \cref{fig:arc_dispersion} at points located on the arc (marked by colored dots in  \cref{fig:fermi_surface}(a)/(b)) for the Hubbard model and \cref{eq:G_our_model}, respectively. 
The momentum cuts start at the node (top), where a single peak, reminiscent of a correlated metal, is visible. This peak becomes broader towards the AFZB, where it starts to split into two, which also marks the onset of the LS. The differences between the D$\Gamma$A and our analytical spectra can be easily rationalized in terms of the simplified interaction in Eq.~(\ref{eq:our_model}), which preserves $\occopp_{\textbf{k}}$ as good quantum number (and, hence, the sharpness of the spectra along the Fermi arcs) and is momentum independent (which reflects into the fixed size of the interaction-driven gap). The overall evolution of the spectral functions, though, as well as of the corresponding Fermi/Luttinger arcs, display an appreciable similarity. 
Here one should note that, in spite of the good agreement of the spectral properties in the temperature range considered, capturing the overall $T$-dependence of the  AF fluctuations in $\dga$ would require a $T$-dependent  $\V(T)$ in  Eq.~(\ref{eq:our_model}).  A fixed, $T$-independent $\V$ would model AF fluctuations essentially
unaffected by temperature.

\noindent
{\sl Conclusions}---We have introduced a model Hamiltonian that, despite being {\sl analytically solvable}, is able to capture, in the temperature range considered, the  doping evolution of the Fermi surface as it is experimentally observed in hole/electron-doped cuprates {\sl and} realized in cutting-edge numerical calculations for the doped Hubbard model, like the D$\Gamma$A employed here.

The advantage of having a full analytical description at our disposal has allowed us to rigorously investigate fundamental questions about the spectral properties of the normal phase of unconventional superconductors.
In particular, our results (i) demonstrate that the emergence of disconnected parts of the FS (``Fermi arcs'') is possible,  (ii)  unveil the specific way of how this can be realized, as well as (iii) capture the differences between electron and hole doping.
In particular, we have shown how Fermi arcs and Luttinger arcs, the latter ones invisible to ARPES, can reside on the same surface and {\sl continually evolve} into one another, lifting the apparent necessity for closed Fermi surfaces and/or Fermi pockets with strongly varying spectral intensity. More precisely, in the presented model, the connection between FS and LS occurs at the $\vk$-points where both cross the AFZB. The Fermi or Luttinger nature of the resulting segments (internal or external w.r.t.~the AFZB) depends on the hole vs.~electron kind of the doping, consistent with the differences of the spectra at the Fermi level characterizing electron- and hole-doped cuprates.

Eventually, we have transparently interpreted the analytically calculated transformation of a Fermi surface into Fermi and Luttinger arcs in terms of the physics of an interaction-driven momentum-selective insulator, arising from the coupling of electrons whose momenta are linked by AF fluctuations. \\

\begin{acknowledgments}
\noindent
{\sl Acknowledgments}---We thank Benjamin Klebel-Knobloch, Juraj Krsnik, Michele Fabrizio, Markus Wallerberger, Thomas Schäfer, Pietro Maria Bonetti, Niklas Witt, Michael Meixner, Mário Malcolms de Oliveira, Giorgio Sangiovanni, and Domenico Di Sante for insightful discussions.  Matthias Reitner acknowledges support as a recipient of a DOC fellowship of the Austrian Academy of Sciences.  We further acknowledge funding through the Austrian Science Funds
(FWF) projects ID I 5398, P 36213, SFB Q-M\&S (FWF project ID F86), and Research Unit
QUAST by the
Deutsche Foschungsgemeinschaft (DFG-project ID FOR 5249, FWF
project ID I 5868). Calculations have been performed on the Vienna Scientific Cluster (VSC).
For the purpose of open access, the authors have applied a CC BY public copyright licence to any Author Accepted Manuscript version arising from this submission.
\end{acknowledgments}
\bibliography{main} 
\end{document}


\author{Paul Worm\,\orcidlink{0000-0003-2575-5058}}
\thanks{These authors contributed equally}
 \affiliation{Institute of Solid State Physics, TU Wien, 1040 Vienna, Austria}
 \author{Matthias Reitner\,\orcidlink{0000-0002-2529-0847}}
\thanks{These authors contributed equally}
 \affiliation{Institute of Solid State Physics, TU Wien, 1040 Vienna, Austria}
 \author{Karsten Held\,\orcidlink{0000-0001-5984-8549}}
 \affiliation{Institute of Solid State Physics, TU Wien, 1040 Vienna, Austria}
  \author{Alessandro Toschi\,\orcidlink{0000-0001-5669-3377}}
 \affiliation{Institute of Solid State Physics, TU Wien, 1040 Vienna, Austria}

\title{Supplementary Material for "Fermi and Luttingers arcs: two concepts, realized on one surface"}

\begin{abstract}

\end{abstract}
\date{\today}

\begin{abstract}
This supplementary material contains additional information to support the conclusions reached in the main text. Section \ref{sec:data} refers to the data availability of the publication. In Section \ref{sec:dga-comp-details}, we outline the computational details for the $\dga$ results. Section \ref{sec:dga-low-T} provides additional $\dga$ results for a lower temperature ($\beta = 22.5/t$). In Section \ref{sec:motivation}, the momentum-selective interaction of our proposed model Hamiltonian is discussed in relation to spin fluctuations of the Hubbard model. Section \ref{sec:luttinger_zeros_a_qualitative_perspective} details a qualitative picture of Luttinger zeros and the analytic structure of the Green's function, given different self-energies. This is connected to the self-energy of the $\dga$ solution for the Hubbard model at the nodal and antinodal points. In Section \ref{sec:high-symmetry-path}, we show that momentum cuts of the spectral function at the node (antinode) show a similar structure as qualitatively discussed in the previous section. Section \ref{sec:pos_t'} discusses the different formation of Fermi arcs for positive or negative values of the hopping parameter $t'$, while  section \ref{sec:loss} treats the implications of the Fermi arcs on the loss of spectral weight and the reduction of the linear specific heat coefficient. In section \ref{sec:highT}, we discuss the specific high-temperature dependence of the model Hamiltonian, which arises due to the intrinsic momentum selectivity and in the last section \ref{sec:V_dep} we compare its spectral function, Fermi, and Luttinger surface in the nodal Fermi arc regime for different values of interaction $\V$.
\end{abstract}

\maketitle

\section{Data availability}
\label{sec:data}
 A data set containing all numerical data and plot scripts used to generate the figures of this publication is publicly available at \cite{data}.

\section{Computational details}
\label{sec:dga-comp-details}
The dynamical mean-field calculations (DMFT), which are used as the starting point for the dynamical vertex approximation ($\dga$), were performed using the continuous-time Quantum Monte Carlo in its hybridization expansion (CT-HYB) implemented in \verb|w2dynamics| \cite{wallenberger2019w2dynamics}. For all quantities worm sampling was used and to ensure high-quality statistics we used about $10^9$ measurements. 
The $\dga$ calculations were performed using the \verb|DGApy| \cite{worm2023} framework. A $\vk$-mesh of (140,140,1) grid points was used, to ensure high-quality momentum resolution. We used $60$ and $80$ positive Matsubara frequencies in the vertex function for $\beta = 12.5/t$ and $22.5/t$, respectively. The asymptotic treatment of the irreducible vertex is described in \cite{kitatani2022} and we used $200$, respectively $250$ additional frequencies for the shell region. The $\lambda$ correction was performed both in the charge and the spin channel \cite{WormThesis, Katanin2009}. 

Analytic continuation was performed using the \verb|ana_cont| package~\cite{Kaufmann2021}. We employed the ''chi2kink'' \cite{Kraberger2017} method for determining the hyperparameter $\alpha$.
To avoid numeric instabilities of sharp delta-peaks we set a minimum imaginary part of the self-energy $\delta = -0.04$, i.e. we set $\Im(\Sigma_{\text{clip}}) = \text{min}(\Im(\Sigma),\delta)$.

\section{$\dga$ at lower temperatures}
\label{sec:dga-low-T}

\begin{figure}[tb]
	\includegraphics[width=1\columnwidth]{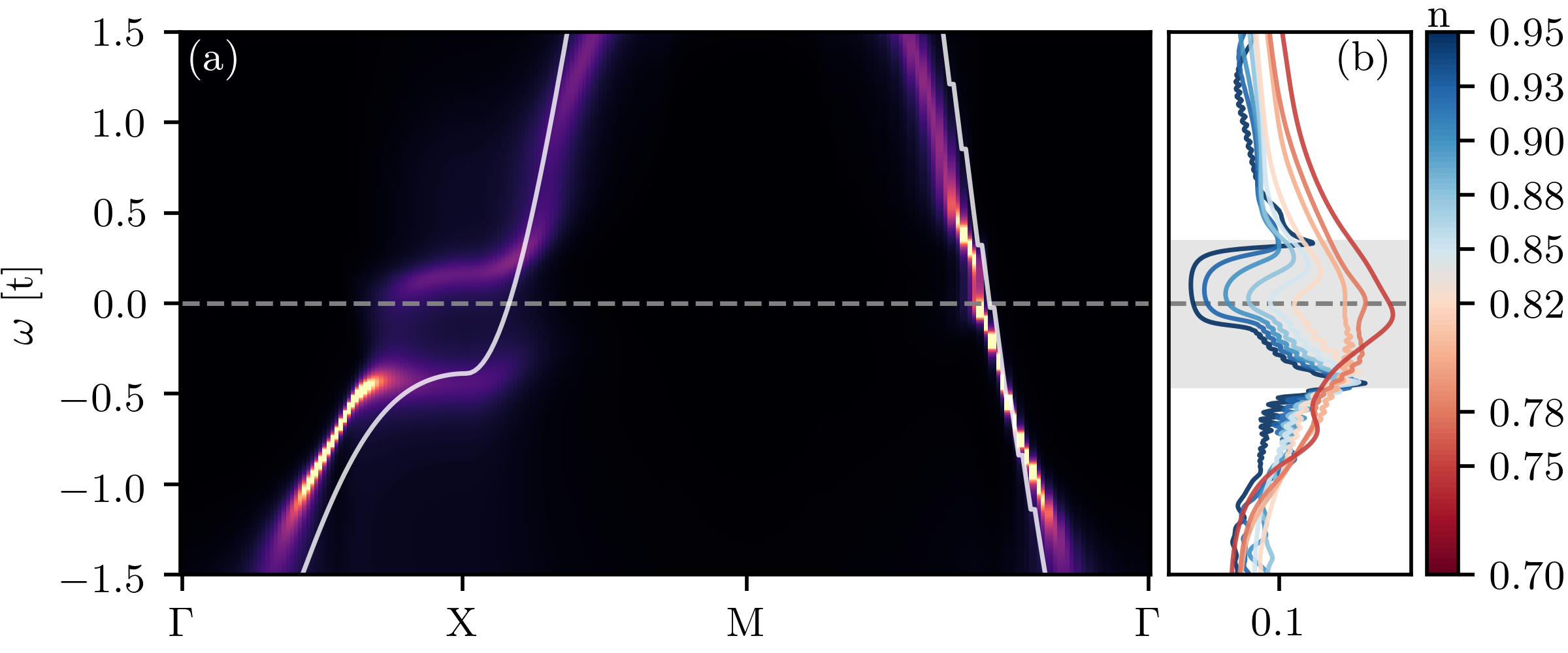}
	\caption{Same as Fig. 2 in the main text, but for $\beta = 22.5/t$. }
	\label{fig:pseudogap_doping_dependence_low_t}
\end{figure}

Below in \cref{fig:pseudogap_doping_dependence_low_t} we show the same plot as in the main text, but for $\beta = 22.5/t$ instead of $\beta = 12.5/t$. As one would expect spectral features like the momentum-selective gap opening around the X point (antinode) are more pronounced. 

\FloatBarrier
\section{Motivation for the momentum-selective interaction}
\label{sec:motivation}

\begin{figure}[tb]
	\includegraphics[width=0.5\columnwidth]{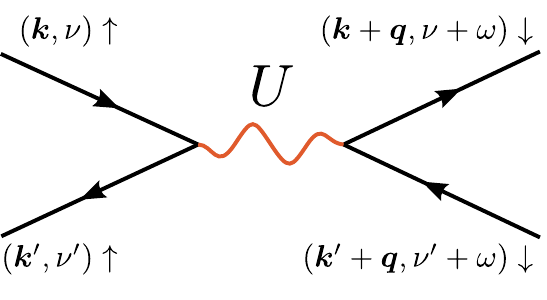}
	\caption{Feynman-diagrammatic representation of the Hubbard interaction in $k$-space. For the proposed model, only $\bm{q}=\bm{Q}=(\pi,\pi)$  and momenta $\bm{k}=\bm{k}'$ contributions are considered, the full frequency dependence of the interaction ($\nu,\nu',\omega$) remains.}
	\label{fig:feyn_u}
\end{figure}

That the interaction term of our model in Eq.~(1) of the main text is able to effectively mimic relevant features of the non-local correlations arising from the Hubbard interaction in the pseudogap regime is far from obvious. In fact, when taking a Fourier transform of the local Hubbard interaction to momentum space (cf. \cref{fig:feyn_u}) one cannot directly see that certain interaction terms are more dominant than others a priori:
\begin{equation}
\frac{U}{2} \sum_{\sigma i} n_{\sigma i} n_{-\sigma i} = \frac{U}{2} \sum_{\vk \vk' \vq} c^\dagger_{\sigma \vk} c^\dg_{\sigma \vk'}  c^\dagger_{-\sigma \vk'+\vq} c^\dg_{-\sigma \vk+\vq}
\end{equation}
(note that we choose here the less used particle-hole transverse momentum notation, which is particular useful in the subsequent discussion). However, by considering the Feynman diagrammatic expansions for the square lattice---due to nesting and, possibly, also due to superexchange processes at intermediate-to-large $U$---we can expect that couplings of the form of $G^0_\vk G^0_{\vk +\vQ} \frac{U}{2} G^0_{\vk'} G^0_{\vk'+\vQ}$, $Q=(\pi, \pi)$, will yield the dominant contributions for low temperatures and near half-filling, especially in ladder diagrams of the spin channel. This can be readily seen, e.g., by performing a random phase approximation (RPA) for the Hubbard model. On the other hand, strong correlation effects are expected to play a key role in the description of the pseudogap and a perturbative method like RPA, where the irreducible vertex is kept fixed to the bare Hubbard interaction, is largely insufficient. The model we propose allows us to take a different approach, by restricting the interaction in momentum-space to $\vq=\vQ$, thereby considering the specific coupling corresponding to the dominant spin-fluctuations. Further, the interaction is restricted to $\delta_{\vk \vk'}$, inspired by the HK model, to make the model exactly solvable while still possibly incorporating strong correlation effects. Previous studies of the HK model demonstrated that interactions diagonal in momentum space, often allow for an analytical study of otherwise hard to tackle many-body physics phenomena, such as Mott like transitions, e.g. in Refs.~\cite{Phillips2020, Wysokinski2023}. In particular, Refs.~\cite{Yang2021exactly,Wang2023} have explored the specific effects of a $\vk$-dependent modification of the HKM interaction.\\

\begin{figure}[tb]
	\includegraphics[width=0.95\columnwidth]{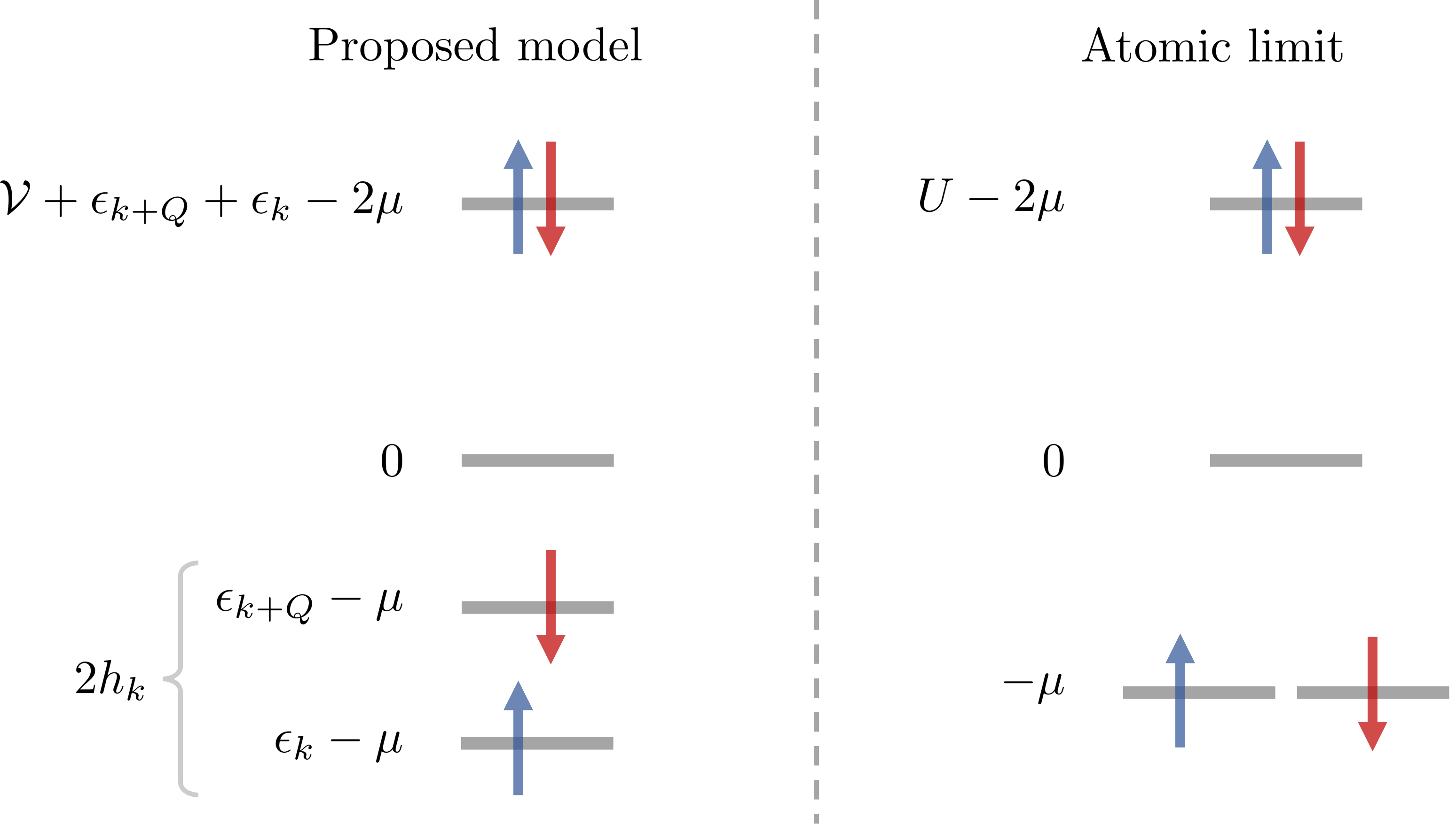}
	\caption{Schematic comparison between the energy levels of the proposed model and the atomic limit. The energy levels of the proposed model (left), illustrate that the coupling between $\vk$, $\uparrow$ and $\vk+\vQ$, $\downarrow$ costs an additional energy $\uafm$ if both momenta are occupied. The Hubbard interaction $U$ (right) affects a double-occupied single site. This sketch refers to the specific situation of $\uafm > \epsilon_{k+Q} + \epsilon_{k} - 2\mu > 0 > \epsilon_{k+Q} + \mu > \epsilon_{k} + \mu$ for the proposed model as well as for $U/2 > \mu > 0$ in the atomic limit of the Hubbard model.} 
	\label{fig:enegry_levels}
\end{figure}

Fig.~\ref{fig:enegry_levels} provides an illustrative comparison of the energy levels coupled by the interaction in the proposed model (left) and the Hubbard model (right).

\FloatBarrier
\section{Luttinger zeros - a qualitative perspective}
\label{sec:luttinger_zeros_a_qualitative_perspective}

\begin{figure}[H]
	\includegraphics[width=0.95\textwidth]{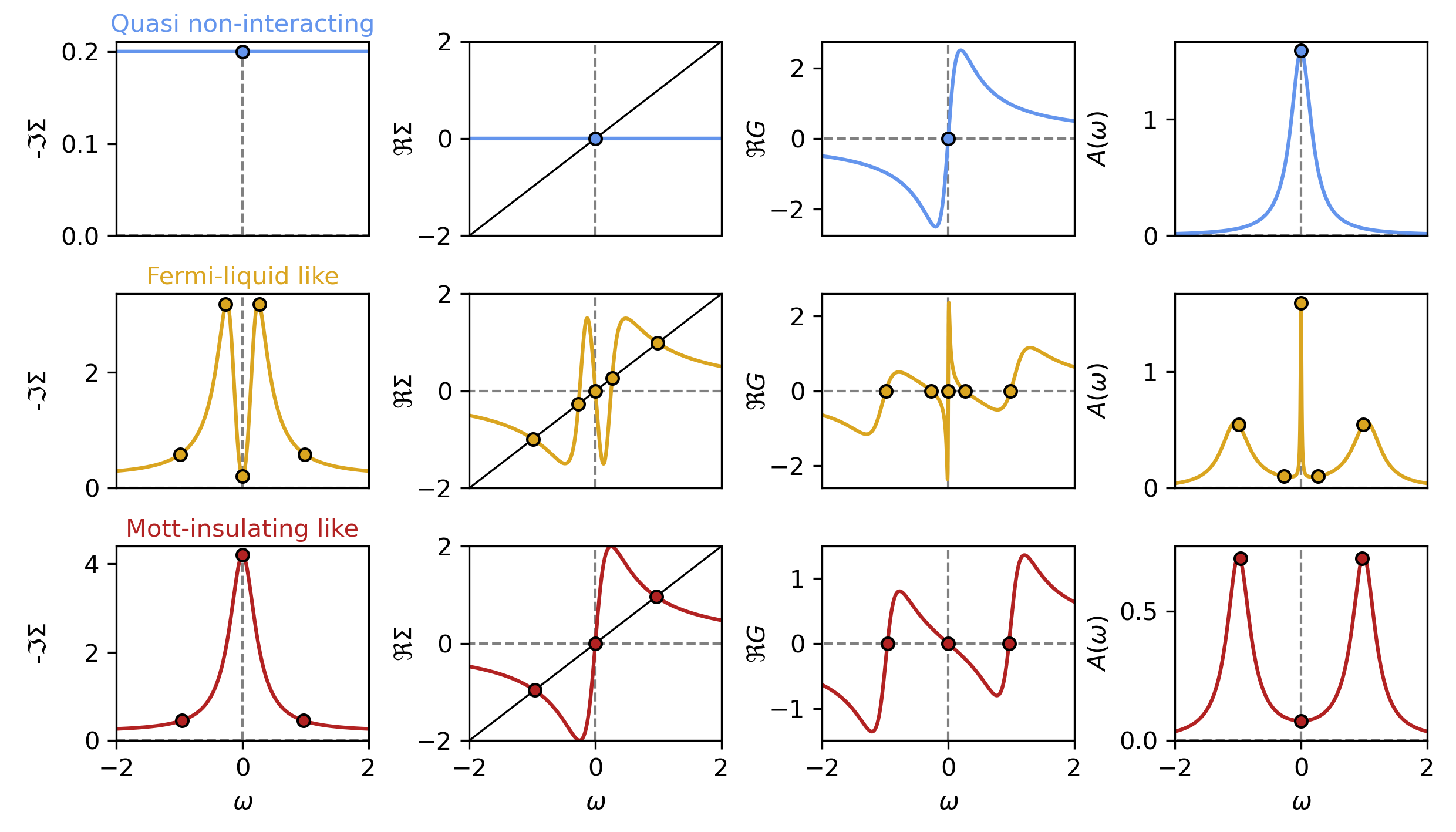}
	\caption{Influence of the functional form of the self-energy on the structure of the Green's function at a Fermi vector $\kf$. From left to right: minus the imaginary part of the self-energy $-\Im \Sigma$, the real part of the self-energy $\Re\Sigma$, the real part of the Green's function $\Re G$, the spectral function $A(\omega)$. 
    From top to bottom: self-energy from \cref{eq:const_scatter} mimicking an effectively non-interacting system, self-energy from \cref{eq:fermi_liquid_scatter} mimicking a renormalized metal, self-energy from \cref{eq:insulator_scatter} mimicking an interaction driven insulator. The black line in the second column corresponds to $\omega - \epsilon(\kf) + \mu$ and the intersection with the real part of the self-energy determines the energy location of the ``bands``. Whether or not these ``bands`` are also visible in the spectral function depends on the magnitude of $\Im \Sigma$.}
	\label{fig:qual_analysis}
\end{figure}

To better understand how zeros of the Green's function arise in the presence of interactions, we discuss three scenarios in \cref{fig:qual_analysis}. Let us note that the discussion below is qualitative in nature and only serves to illustrate how different functional forms of the self-energy $\Sigma$ change the spectral function $A(\omega)$, which is accessible in ARPES experiments. Furthermore, we discuss the appearance of zeros in the Green's function. 

The single-particle Green's function is given by 

\begin{equation}
    G(\omega,\textbf{k}) = \frac{1}{\omega - \epsilon(\textbf{k}) + \mu - \Sigma(\omega,\textbf{k})},
\end{equation}
where $\epsilon(\textbf{k})$ is the energy-momentum dispersion and $\mu$ the chemical potential. For the following, we consider the Green's function at a single Fermi vector $\kf$, defined as a momentum vector that satisfies

\begin{equation}
    \omega -\epsilon(\kf)+\mu-\Re \Sigma(\omega,\kf) = 0,
\label{eq:band_equation}
\end{equation}
at $\omega = 0$.

To best understand the influence of the self-energy on the Green's function, we plot the corresponding (minus) imaginary part of the self-energy ($-\Im \Sigma$) in the first row, the real part ($\Re \Sigma$) in the second, the real part of the Green's function ($\Re G$) in the third and finally the spectral function ($A(\omega) = -\frac{1}{\pi} \Im G$) in the rightmost column. 
 
In the first scenario (first row in blue), the scattering rate is constant ($\Delta_{0}$) with a high-frequency cut-off
\begin{equation}
    \Im \Sigma = - \Delta_{0} \enskip \Theta(\omega^{*}-|\omega|),
\label{eq:const_scatter}
\end{equation}
where $\Delta_0$ is a positive real constant and $\omega^{*}$ is some frequency larger than the bandwidth. This essentially describes the \emph{quasi} non-interacting case with some finite lifetime of the states, e.g. due to impurity scattering \footnote{The physical origin of the self-energy is not relevant for the qualitative discussion here.}. The resulting spectrum is a single Lorentzian peak centered around the Fermi energy (here $\omega = 0$), where the real part of the Green's function changes sign (marked by a dot). 

For the second case, the self-energy is inspired by that of Fermi-liquids, or renormalized metals, 

\begin{equation}
    \Im \Sigma = - \Delta_{1} \enskip \frac{1+(\sigma \omega)^2}{1+(\sigma \omega)^4} - \Delta_{0} \enskip \Theta(\omega^{*}-|\omega|),
\label{eq:fermi_liquid_scatter}
\end{equation}
with $\sigma$, $\Delta_0$, and $\Delta_1$ positive real constants.
In addition to the constant scattering rate of \cref{eq:const_scatter} a term which scales as $\omega^2$ for small frequencies, reminiscent of Fermi-liquids, has been added. The $1/\omega^4$ ensures the correct high-frequency asymptotics of the self-energy and the real part has been computed using the Kramers-Kronig relations \cite{Kramers1927,Kronig1926}. 

The resulting spectrum, center row (orange) in Fig.~\ref{fig:qual_analysis}, displays a three-peak structure, similar to DMFT solutions for the correlated metallic regime of the square-lattice Hubbard model, \cite{Qin2022,WormThesis}. Already this simple self-energy is enough for the appearance of zeros in the Green's function, which separate the three ``poles''. Both poles and zeros correspond to solutions of \cref{eq:band_equation}, which is indicated by the intersection of the black line ($\omega - \epsilon(\kf) + \mu$) and $\Re \Sigma$ (colored line). The difference (w.r.t.~finite background scattering) is that the poles correspond to a small $\Im \Sigma$ and large spectral weight, similar to the quasiparticle and Hubbard side-bands in DMFT. At the zeros of the Green's function, however, spectral weight is notably absent and they are consequently invisible in ARPES.

For the ``Fermi-liquid-like'' case these zeros are located away from the Fermi energy. However, in interaction-driven insulators, e.g. Mott insulators, these zeros can be directly at the Fermi energy \cite{Stanescu2007}. \cref{eq:insulator_scatter} displays a simple self-energy, which replicates this behavior, 

\begin{equation}
    \Im \Sigma = - \Delta_{1} \enskip \frac{1}{1+(\sigma \omega)^2} - \Delta_{0} \enskip \Theta(\omega^{*}-|\omega|),
\label{eq:insulator_scatter}
\end{equation}
where the crucial part is that the self-energy has a peak at (or close to) the Fermi energy. This suppresses the spectral weight at the Fermi energy, which becomes a Luttinger zero. At the same time, the corresponding (frequency dependent) real part of the self-energy creates two additional solutions for \cref{eq:band_equation}, which correspond to the two peaks visible in the spectrum above and below the Fermi energy.

This qualitative picture outlines what is happening to the Fermi surface in the pseudogap: In the hole-doped case, the self-energy at the node is qualitatively similar to the ``Fermi-liquid-like'' one discussed here and the corresponding spectrum appears metallic, yielding the Fermi arcs. At the antinode, however, the self-energy developes a peak close to the Fermi surface, resulting in a gapped spectrum with a Luttinger zero: the Luttinger arc, which is invisible in ARPES. For the electron-doped case, this picture is simply reversed.

\begin{figure}[H]
	\includegraphics[width=0.95\textwidth]{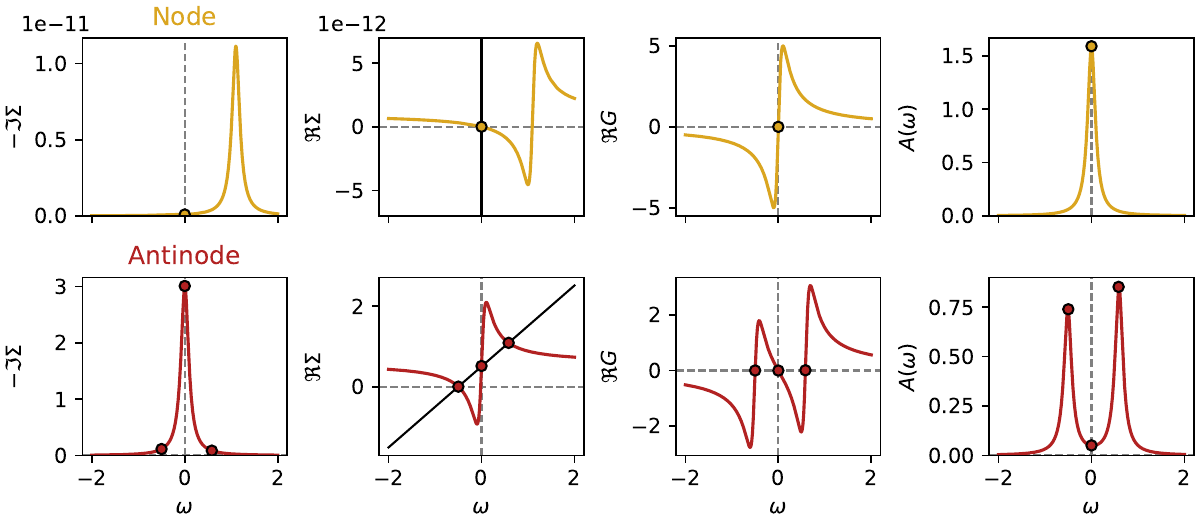}
	\caption{Similar to Fig.~\ref{fig:qual_analysis}, showing the difference of the functional form between solutions on the node (Fermi arc) and antinode (Luttinger arc) for the proposed model in the nodal Fermi arc regime at $\V= 1.1t$, $n=0.85$, $\beta=12.5/t$, on the real frequency axis, where $\ii\nu_n \to \omega + \ii \delta$ with spectral broadening $\delta=0.1$. From left to right: minus the imaginary part of the self-energy $-\Im \Sigma$, the real part of the self-energy $\Re\Sigma$, the real part of the Green's function $\Re G$, and the spectral function $A(\omega)$. 
    Top: self-energy at the node showing an almost non-interacting solution. Bottom: self-energy at the antinode resembling an interaction-driven insulator. }
	\label{fig:arc_analysis}
\end{figure}

In the following, we compare this qualitative picture to the results of the model Hamiltonian in Eq.~(1) of the main text and to the $\dga$ solution of the Hubbard model, both for the nodal Fermi arc at hole doping.
For the model in Eq.~(1) the situation  for $\V= t$, $n=0.85$, $\beta=12.5/t$ is shown in Fig.~\ref{fig:arc_analysis}. For the plot on the real frequency axis, where $\ii\nu_n \to \omega + \ii \delta$, we use a spectral broadening of $\delta=0.1$. Here, at the node (top row), the vanishingly small self-energy leaves the corresponding spectral function almost unaltered w.r.t. the non-interacting solution, reflecting the fact that the model preserves $\occopp_{\textbf{k}}$ as a good quantum number. As a matter of course, there are no Hubbard bands.  In contrast, for the antinode (bottom row), the self-energy clearly displays the spectral behavior of an interaction-driven insulator.

\begin{figure}[H]
	\includegraphics[width=0.95\textwidth]{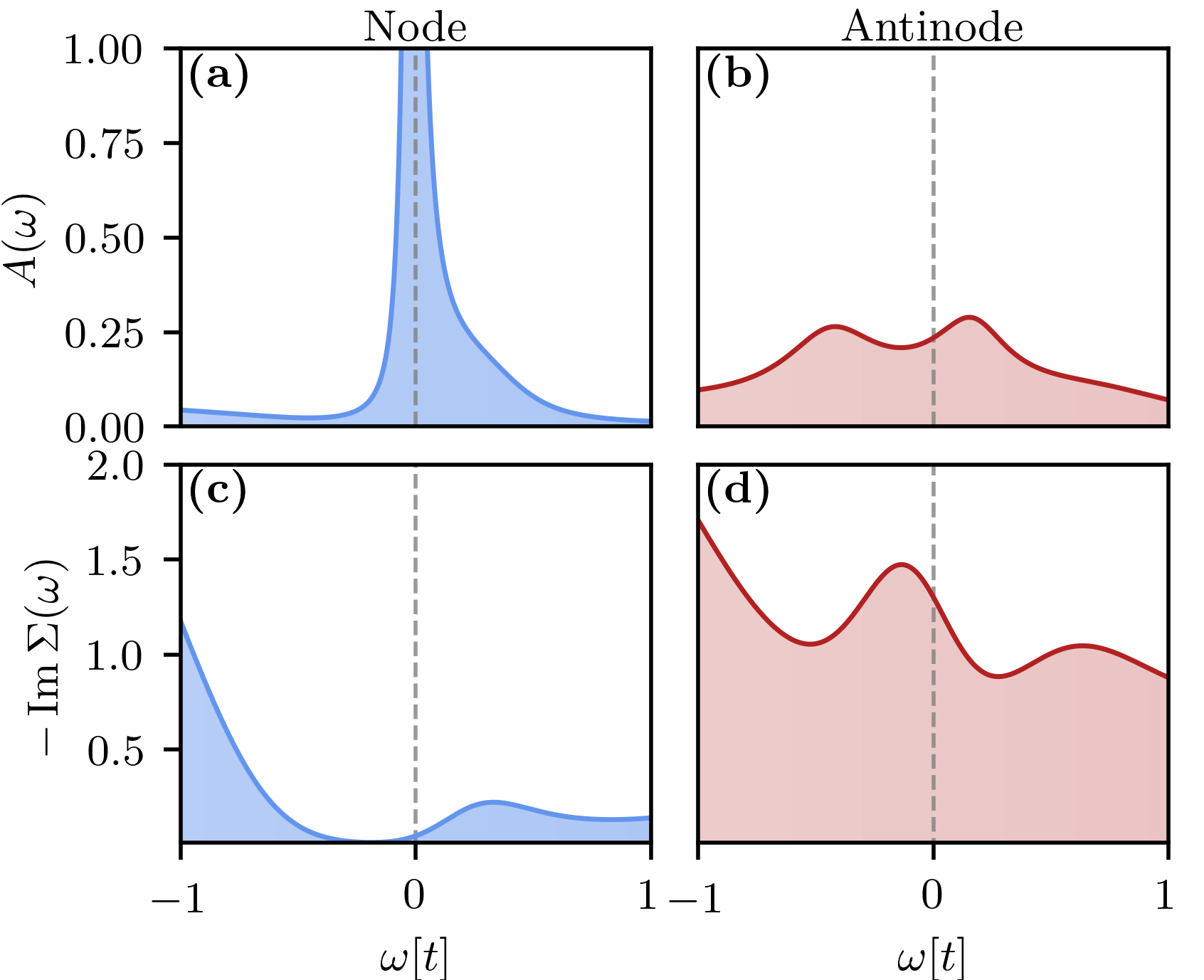}
	\caption{Spectrum (top) and self-energy (bottom) at the node (left) and antinode (right) for the $\dga$ solution of the Hubbard model at $n = 0.85$, $\beta = 12.5/t$ and $U = 8t$.}
	\label{fig:not_qual_analysis}
\end{figure}

Fig.~\ref{fig:not_qual_analysis} displays this behavior for the $\dga$ solution of the Hubbard model at $n = 0.85$ and $\beta = 12.5/t$. For the node (left in blue) the self-energy (b) shows a trough at the Fermi energy ($\omega = 0$), which results in a metallic spectrum (a). On the contrary, the self-energy at the antinode develops a peak close to the Fermi energy (d), which results in a two-peak structure in the spectrum (b).

\FloatBarrier
\section{Spectrum along a high-symmetry path in the Brillouin zone}
\label{sec:high-symmetry-path}

As supplemental information to Fig.~3 in the main text, we plot the spectral function for the same parameters along a high-symmetry path ($\Gamma$-X-M-$\Gamma$) below. The left side shows the $\dga$ solution of the Hubbard model and the right side the exact solution of the proposed model. The lower panels show the spectral function and the real part of the single-particle Green's function along cuts at the node and antinode respectively. 

The momentum-selective insulating behavior in the pseudogap is clearly visible as a double-peak (gapped) structure at the antinode, while the node displays a single peak corresponding to the Fermi arc. Furthermore, the real part of the Green's function contains information about the location of the Luttinger surface, as a zero-crossing without spectral weight and is observed within the gap. 

\begin{figure*}[tb]
	\includegraphics[width=0.95\textwidth]{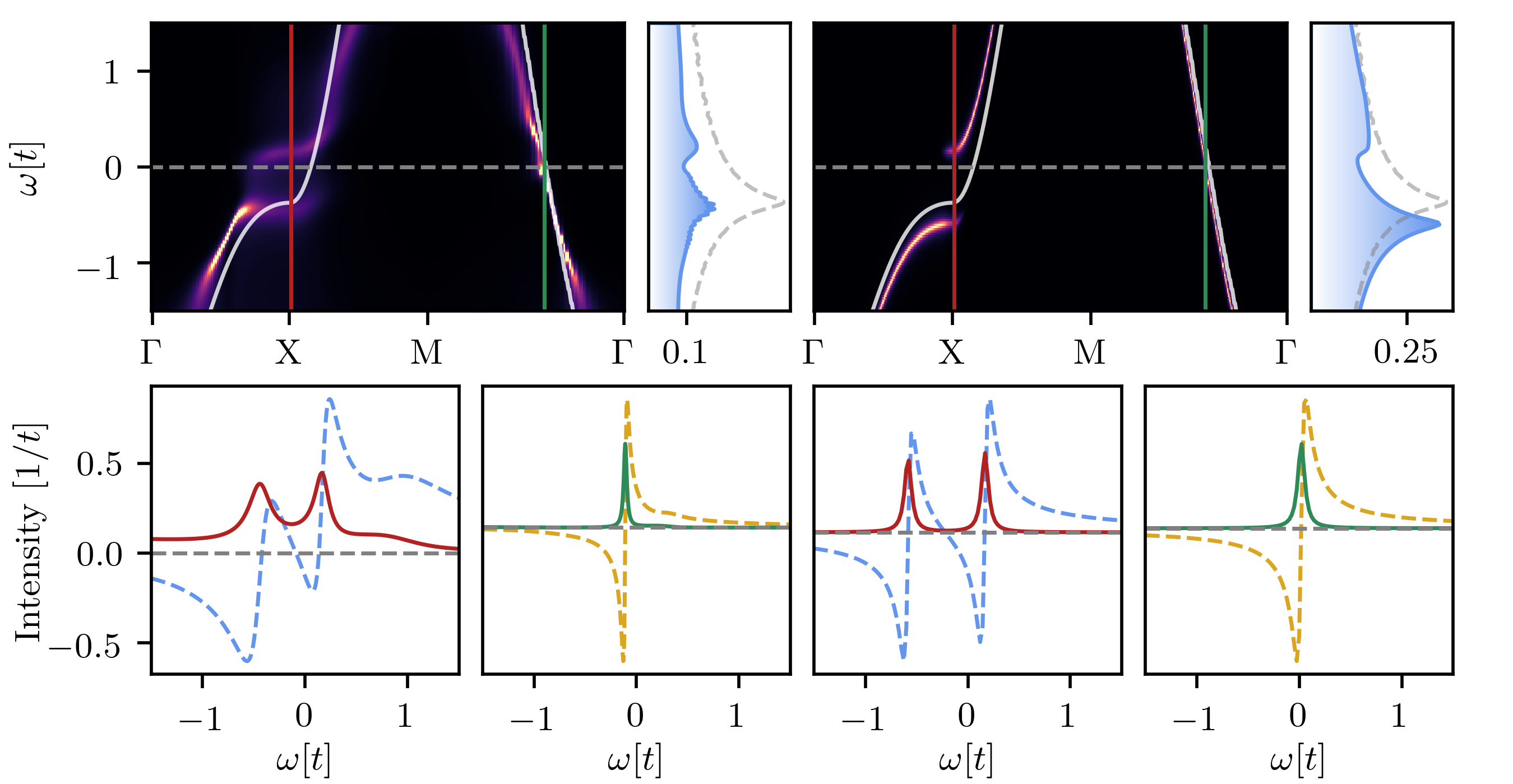}
	\caption{Top row: $A(\omega,k)$ along a high-symmetry path, $\Gamma = (0,0)$, X$ = (\pi,0)$ and M$ = (\pi,\pi)$, in the Brillouin zone for $n = 0.85$, $\beta = 22.5/t$, and corresponding momentum-integrated spectral function.  The white lines display the non-interacting band dispersion for the same filling and the dashed gray lines show the corresponding non-interacting momentum-integrated spectral function.  Left: $\dga$ solution of the Hubbard model at $U = 8t$. Right: analytic solution of the proposed model at $\V = 1.1t$.  Lower panels below the top panels display respective momentum cuts of the spectral function (solid) and real part of the Green's function (dashed) for two momenta at the X point (left) and the node (right).}
	\label{fig:symmetry_path}
\end{figure*}

\FloatBarrier
\section{Fermi and Luttinger arcs for positive $t'$}
\label{sec:pos_t'}

\begin{figure}[t]
	\includegraphics[width=0.6\columnwidth]{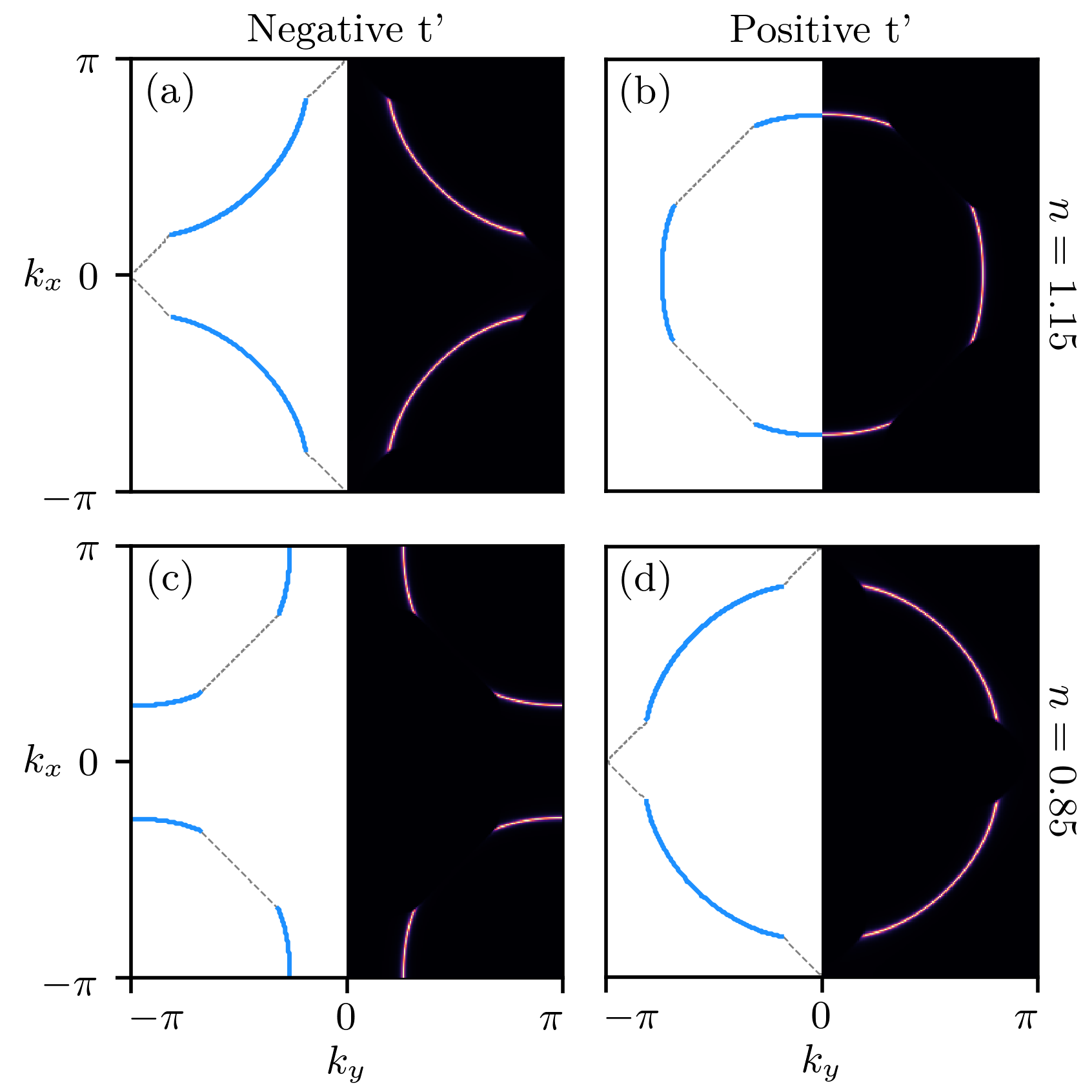}
	\caption{Fermi surface (blue line) and Luttinger surfaces (gray line) computed for the proposed model at hole- (top row) and electron-doping (bottom row) at $\beta = 50/t$ and $\V = 1.1t$. Left column: $t=1$, $t'=-0.2$ and $t'' = +0.1$. Right column: $t=1$, $t'=+0.2$ and $t'' = -0.1$  The left half of the Brillouin zone displays the Fermi surface (blue line) and Luttinger surface (gray dashed), while the right half shows the spectral function at zero frequency ($A(\omega=0)$).}
	\label{fig:fermi_surface_positive_tp}
\end{figure}

In the main text we discussed the spectral function and Fermi/Luttinger arc properties for the tight-binding (tb) parameter set $t=1$, $t'=-0.2$ and $t'' = 0.1$, inspired by Wannier models for cuprates \cite{WormThesis} and previous studies \cite{Nicoletti2010a, krien2022pseudogap}. However, our general description of the mechanism and formation of Fermi and Luttinger arcs is valid for any set of tb parameters we tested. Below in \cref{fig:fermi_surface_positive_tp}, we illustrate this for $t'=+0.2$ and $t'' = -0.1$, i.e. inverting the sign of the next and next-next nearest neighbor hopping. Our results show again Fermi and Luttinger arcs, with the observation of Fermi arcs at the antinode (node), for electron- (hole-) doping. The remainder of the tight-binding Fermi surface becomes gapped and transforms into a Luttinger arc. 

This also illustrates, that for our proposed model and differently to DCA calculations for the Hubbard model \cite{Wu2018}, the existence of a pseudogap does not appear to be affected by the electron/hole nature of the underlying tb FS. That is, in our proposed model, a pseudogap can form for both electron- and hole-like tb parameters, provided that the interaction strength is large enough and the tb FS crosses the AFZB. The respective location of the Fermi and Luttinger arcs in momentum space depends on both the sign of $t'$ and whether the system is hole or electron doped. We summarize this briefly in Table~\ref{tab:Luttinger_arc_loc} below.

\begin{table}[]
    \caption{Location of the Luttinger arc (pseudogap) in spectrum of the proposed model for different signs of $t'$ and dopings.}
\begin{tabular}{c|c|c}
\hline
\hline
 - & negative $t'$ ($t' < 0$)& positive $t'$ ($t' > 0$)   \\ \hline
 hole doping &  \color{red}antinode &  \color{blue}node  \\ \hline
 electron doping &  \color{blue}node&  \color{red}antinode\\ 
 \hline
\hline
\end{tabular}
\label{tab:Luttinger_arc_loc}
\end{table}

On the other end (at least for $t'' \! = \! 0$), we note that the occurrence of the pseudogap is directly associated with a change in the nature of the tb FS. This is, because such change is connected to the tb FS crossing of the AFZB. As a result, for $t'>0$ a pseudogap appears only for electron-like FSs, while for $t'<0$ only for hole-like ones.

\FloatBarrier
\section{Loss of spectral weight and implications for the specific heat}
\label{sec:loss}
\begin{figure}[tb]
    \includegraphics[width=1.\columnwidth]{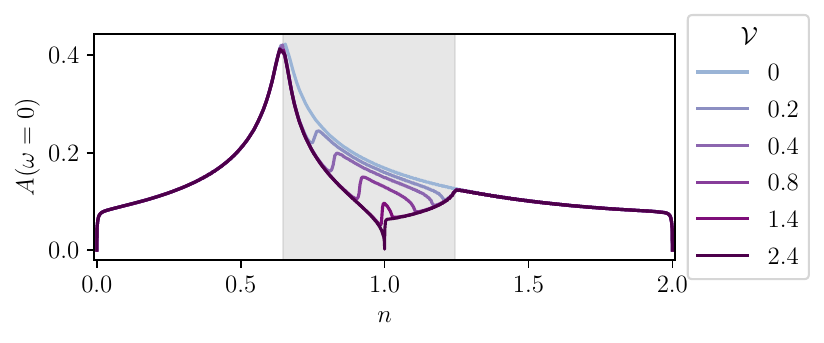}
	\caption{Spectral weight at the Fermi level $A(\omega=0)$ for different interaction strength and electron density for $T\to0$ ($1/T=10^6/t$). In comparison with $\V=0$, the disconnected Fermi surface (gray-shaded area) leads to a depletion of spectral weight. }
	\label{fig:spectral_weight_fermi}
\end{figure}
The model Hamiltonian in Eq.~(1) of the main text allows for a transparent view on the evolution of the disconnected Fermi surfaces  (Fermi arcs)  as function of doping. In \cref{fig:spectral_weight_fermi} the spectral weight at the Fermi level
\begin{equation}
A(\omega=0) = 2 \sum_\vk \left[ (1-n_{\vk+\vQ}) \delta(\xi_\vk) +  n_{\vk+\vQ}\delta(\xi_\vk +\V) \right]
\end{equation}
as function of the electron density $n$ for different interaction strength $\V$  is displayed for $T\to 0$. For the regime, where the Fermi surface (FS) becomes disconnected and evolves into a Luttinger surface (LS) (gray-shaded area) the reduced FS leads to a corresponding loss of spectral weight at the Fermi level in comparison to the non-interacting $\V=0$ case.

In our model, the major loss of spectral weight at the Fermi level is directly reflecting the reduction of the FS (i.e., the length of the Fermi arc). In the low-$T$ behavior of the electronic specific heat $c_V = \frac{\partial \expec{H}}{\partial T} $ a similar effect can be observed, where
\begin{equation}
\label{eq:mean_H}
    \begin{split}
    \expec{H} = &\sum_\vk \frac{1}{Z_\vk}\left[ 2 \xi_\vk\, \e^{-\beta \xi_\vk} + (\xi_\vk + \xi_{\vk+\vQ}+\V)\, \e^{-\beta (\xi_\vk +\xi_{\vk+\vQ}+\V)}\right] \\
    = & \sum_\vk \left[ 2 \xi_\vk   f_{\xi_\vk} (1 - n_{\vk+\vQ})+  2 (\xi_\vk +\V)  f_{\xi_\vk+\V} n_{\vk+\vQ}- \V d_\vk\right],
    \end{split}
\end{equation}
 $Z_\vk = 1 + \e^{-\beta \xi_\vk}+ \e^{-\beta\xi_{\vk+\vQ}}+ \e^{-\beta (\xi_\vk+ \xi_{\vk+\vQ}+\V)}$, $n_{\vk+\vQ}=\frac{\e^{-\beta\xi_{\vk+\vQ}}+ \e^{-\beta (\xi_\vk+ \xi_{\vk+\vQ}+\V)}}{Z_k}$, $d_\vk =\frac{ \e^{-\beta (\xi_\vk+ \xi_{\vk+\vQ}+\V)}}{Z_k}$ and  $f_\xi=\frac{1}{\e^{\beta \xi}+1}$ is the Fermi function.

\begin{figure}[tb]
    \includegraphics[width=1.\columnwidth]{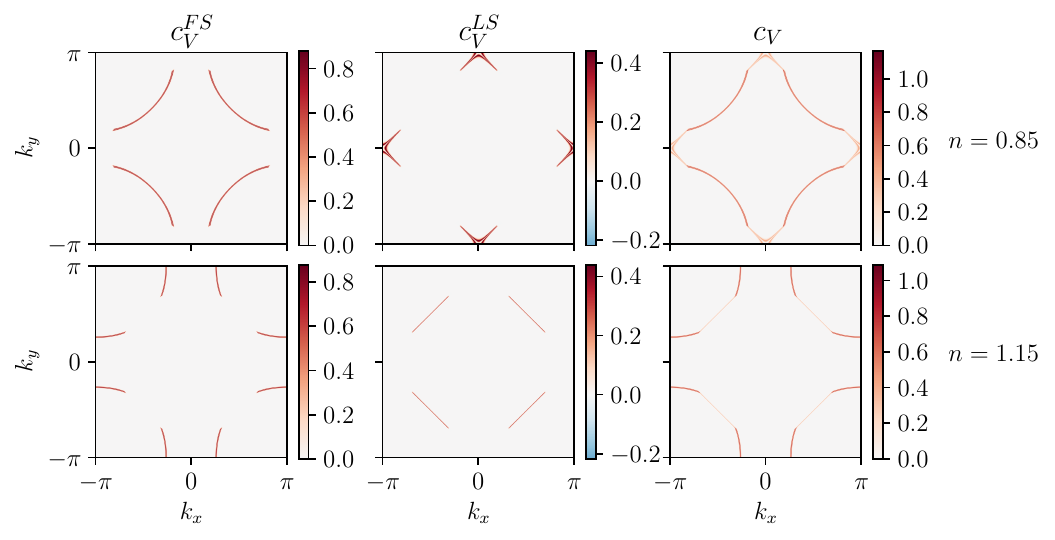}
	\caption{$k$-resolved contributions to the electronic specific heat $c_V=c_V^{FS}+c_V^{LS}$ (right column)  divided into the Fermi surface $c_V^{FS}$ (left column) and Luttinger surface $c_V^{LS}$ (middle column) contributions for hole (top row) and electron (bottom row) doping at interaction strength $\V=2.4 t$ and temperature $T=1/100\, t$.}
	\label{fig:cv_k}
\end{figure}

\begin{figure}[tb]
    \includegraphics[width=1.\columnwidth]{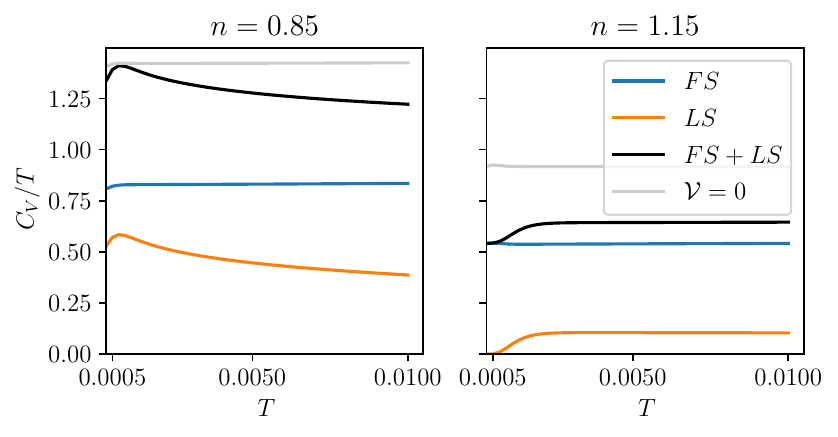}
	\caption{Temperature dependence of the different contributions to the electronic specific heat $c_V/T=c_V^{FS}/T+c_V^{LS}/T$ separated into the Fermi surface $FS$ and Luttinger surface $LS$ contributions for hole (left panel) and electron (right panel) doping with interaction strength $\V=2.4 t$. The specific heat of the corresponding non-interacting model ($\V=0$) is shown as a comparison.}
	\label{fig:cv_T}
\end{figure}

 Taking the derivative of \cref{eq:mean_H} with respect to the temperature $T$, we get
 \begin{equation}
 \label{eq:specific_heat}
     \begin{split}
         c_V = & \overbrace{-2 \beta \sum_\vk \left[\xi_\vk^2 f'_{\xi_\vk} (1 - n_{\vk+\vQ}) +  (\xi_\vk +\V)^2  f'_{\xi_\vk+\V} n_{\vk+\vQ} \right]}^{c^{FS}_V}\\
         &\underbrace{-\beta^2 \sum_k\left[2 \frac{\partial n_{\vk+\vQ}}{\partial \beta} (-\xi_\vk f_{\xi_\vk}+ (\xi_\vk +\V)  f_{\xi_\vk+\V}) -\V \frac{\partial d_\vk}{\partial \beta}\right]}_{c^{LS}_V},
     \end{split}
 \end{equation}
 where the first term on the right hand side of \cref{eq:specific_heat} entails derivatives of the Fermi function and gives the usual contribution of the FS ($c^{FS}_V\propto \gamma T$ for $T\to 0$ ) while the second term yields a ``thermally activated" behavior associated to the LS ($c^{LS}_V$), with
 \begin{equation}
     \begin{split}
         c^{LS}_V = -\beta^2 \sum_\vk &\left[\V^2 d_\vk (d_\vk+1 - f_{\xi_\vk+\V}-f_{\xi_{\vk+\vQ}+\V}) \right.\\
         &+ 2\V d_\vk \left((1-n_{\vk+\vQ})\xi_\vk  (  f_{\xi_\vk} -  f_{\xi_\vk+\V})  + (1-n_\vk)\xi_{\vk+\vQ}  (  f_{\xi_{\vk+\vQ}} -  f_{\xi_{\vk+\vQ}+\V}) \right)\\
         &+  n_{\vk+\vQ}(1-n_{\vk+\vQ}) ( f_{\xi_\vk} - f_{\xi_\vk+\V})  ( \xi_\vk \xi_{\vk+\vQ} -\xi_\vk^2 ( f_{\xi_\vk} - f_{\xi_\vk+\V})  ) \\
         &+ \left. n_\vk(1-n_\vk) ( f_{\xi_{\vk+\vQ}} - f_{\xi_{\vk+\vQ}+\V})  ( \xi_\vk \xi_{\vk+\vQ} -\xi_{\vk+\vQ}^2 ( f_{\xi_{\vk+\vQ}} - f_{\xi_{\vk+\vQ}+\V})  )\right]
     \end{split}
 \end{equation}
 (cf. \cref{fig:cv_k}).
 Hence, in the regime of the disconnected FS, the reduced length of the Fermi arc leads to an effectively {\sl reduced} coefficient $\gamma$ in the specific heat $c_V(T)$ (see \cref{fig:cv_T}).

\FloatBarrier
\section{High Temperatures}
\label{sec:highT}
 
\begin{figure}[t]
	\includegraphics[width=0.9\columnwidth]{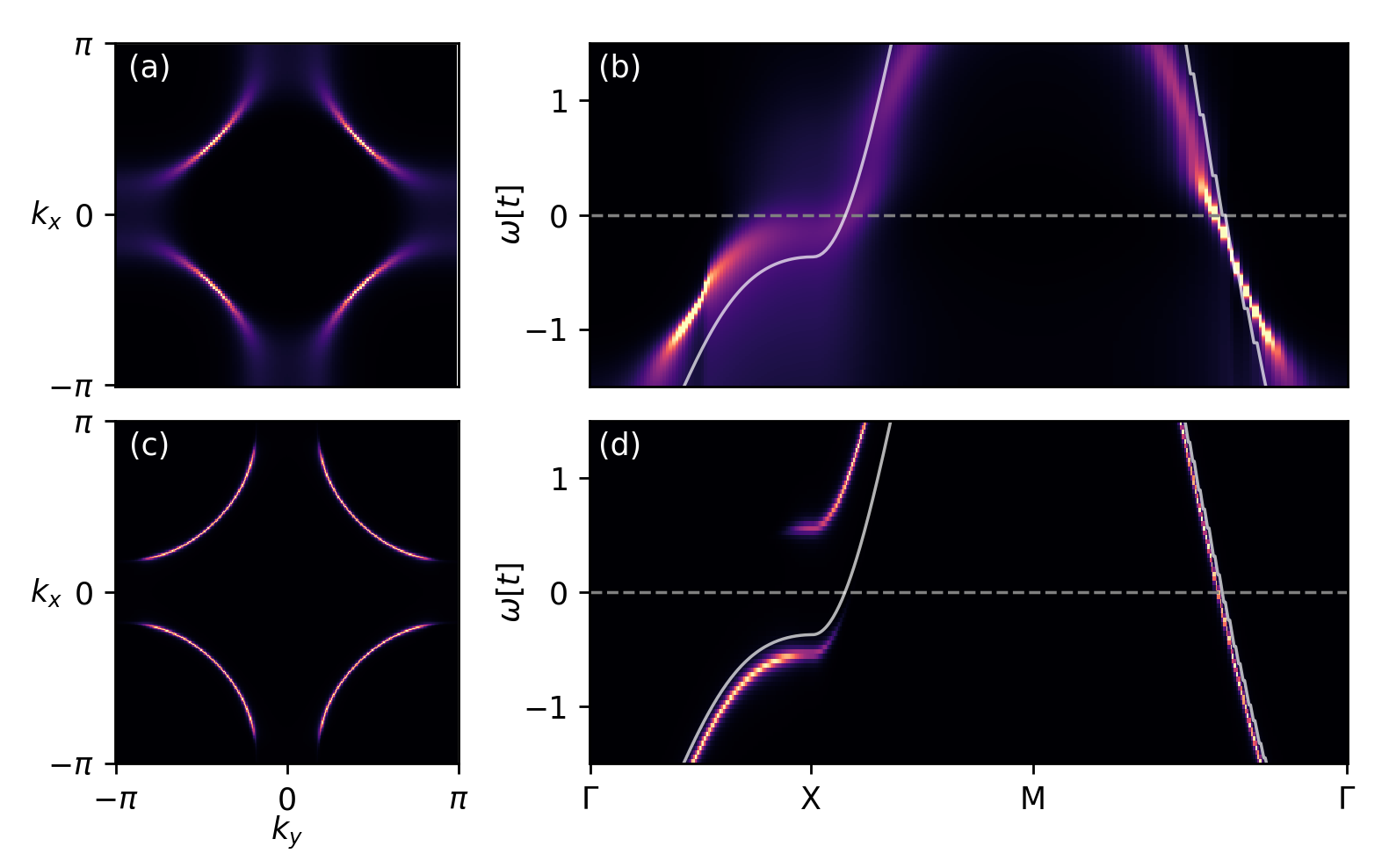}
	\caption{(a,c) Comparison of the high-temperature spectral function at zero frequency $A(\omega = 0,k)$; and  (c,d) of the spectral function $A(\omega,k)$ along the high-symmetry path, $\Gamma = (0,0)$, X$ = (\pi,0)$ and M$ = (\pi,\pi)$, in the Brillouin zone for $n = 0.85$  between $\dga$ at $U=8t$  (top row) and the model Hamiltonian at $\V=1.1t$ (bottom row). Temperature is $\beta=7.5t$ which is somewhat above the pseudogap temperature, where there is a stronger broadening but not a gap opening at the antinode in D$\Gamma$A.} 
	\label{fig:dispersion_high_T}
\end{figure}

As mentioned right before the Conclusion in the main text, if one keeps the interaction strength $\V$ fixed to an appropriate value (e.g., in our case, $\V=1.1t$) that describes the AF correlations in the pseudogap regime of the Hubbard model, it cannot be expected that the overall good agreement with the $\dga$ spectral functions found at low-to-intermediate temperatures also holds in the high-$T$ regime. 
We exemplify this effect by showing the comparison of the $\dga$ ($U= 8t$) and model ($\V=1.1 t$) spectral functions at $\beta = 7.5/t$ in \cref{fig:dispersion_high_T}: Here we observe that 
while the momentum-selective effects of $\dga$, driven by nesting and AF spin fluctuations become gradually suppressed at higher temperatures, the built-in momentum selectivity of the model Hamiltonian represents an intrinsic property of its physics. In particular, a sharp gap remains always present in the spectral function of the model Hamiltonian, while the two corresponding ``bands" become gradually more filled with higher temperatures. 
In order to describe the temperature dependence of the spectra of the Hubbard model more accurately, one could introduce a temperature-dependent interaction $\V(T)$ which decreases with temperature, becoming negligible above the pseudogap regime.

\section{Interaction dependence of the model Hamiltonian}
\label{sec:V_dep}

\begin{figure}[t]
	\includegraphics[width=0.9\columnwidth]{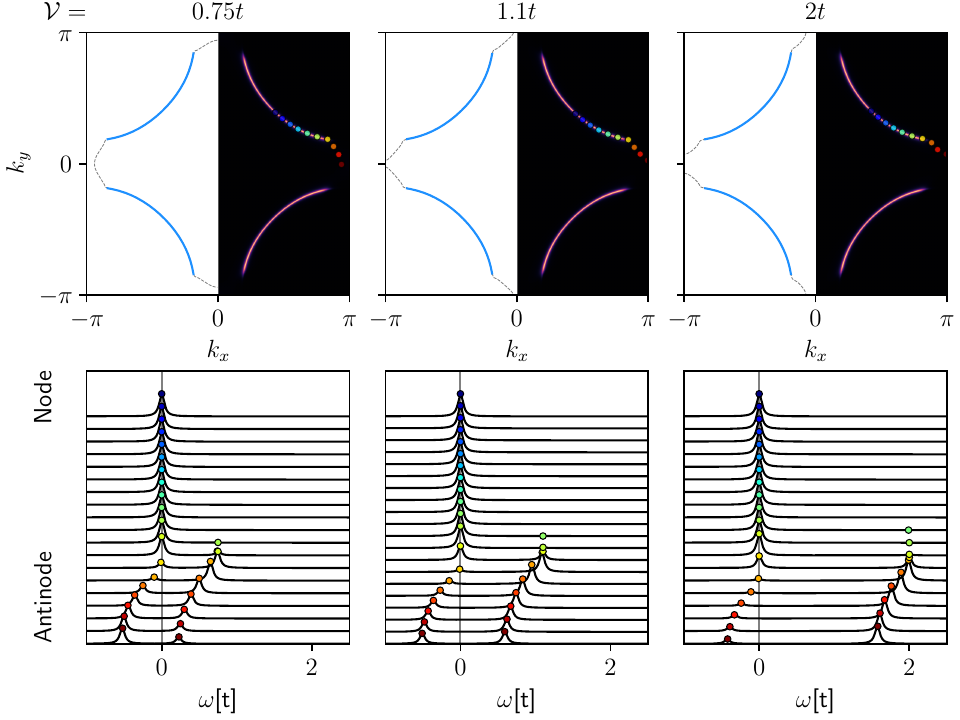}
	\caption{Dependence on the model interaction value, $\V$,  of the results Fig.~2 and Fig.~4 of the main text. 
  Columns from left to right: $\V=0.75t,\, 1.1t,\, 2t$; $\beta=12.5/t$ and $n=0.85$.
  Top row: Fermi (blue line), and Luttinger surfaces (gray dashed line) in the left half of the  Brillouin zone; spectral function at zero frequency $A(\omega = 0,k)$  in the right half of the  Brillouin zone. Colored dots mark the locations for the cuts in the spectral function below. Bottom row: $A(\omega,k)$ cuts along the Fermi-Luttinger surface for the momenta indicated by the color above.}
	\label{fig:V_comparision}
\end{figure}

In Fig.~\ref{fig:V_comparision} we display the interaction $\V$ dependence of the spectral function, Fermi and Luttinger surfaces for the model Hamiltonian in the nodal Fermi arc regime (cf. Fig.~1 in the main text) at inverse temperature $\beta=12.5/t$ and $n=0.85$. For the $A(\omega,k)$ cuts along the Fermi-Luttinger surface (bottom row panels),  at the Luttinger surface (gaped spectrum), the position in $\omega$ of the lower peaks is almost unaffected by an increase in $\V$, though its spectral weight gets reduced. The position of the higher second peak, on the contrary, gets shifted to higher values of $\omega$ with increasing spectral weight. This shift of spectral weight at the Luttinger surface results in a slightly reduced length of the Fermi arc for increasing values of $\V$ (top row panels).

\FloatBarrier

\bibliography{main}